\documentclass[aps,pra,preprint,groupedaddress]{revtex4-2}
\usepackage[top=2.5cm,bottom=2.5cm,right=3cm,left=3cm]{geometry}
\usepackage[utf8]{inputenc}
\usepackage{braket}
\usepackage{amsmath}
\usepackage{bbm}
\usepackage{appendix}
\usepackage{amsfonts}
\usepackage{comment}
\usepackage{cancel}
\usepackage{bbold}
\usepackage{graphicx}
\usepackage{float}
\usepackage{MnSymbol}
\usepackage{mathtools}
\usepackage{soul}
\usepackage{tabularx}
\usepackage{array}
\usepackage{accents}
\usepackage{bm}
\usepackage{placeins}
\usepackage{mathrsfs}

\newcolumntype{Y}{>{\centering\arraybackslash}X}

\usepackage[colorlinks, linkcolor=blue]{hyperref}
\usepackage[nameinlink, capitalise]{cleveref}

\newcounter{definition}
\newtheorem{Definition}[definition]{Definition}

\newcounter{theorem}
\newcounter{proposition}
\newtheorem{Proposition}[proposition]{Proposition}

\newcommand{\domark}{%
	\vbox to 0pt{
		\kern-\dp\strutbox
		\hbox{\smash{\llap{*\kern1em}}}
		\vss
	}%
}

\begin{document}

	\title{Quantum-like states from classical systems}
	\author{Gregory D. Scholes}
	\address{Department of Chemistry, Princeton University, Princeton, NJ 08544, USA}
	\email{gscholes@princeton.edu}

	\date{\today}

	\begin{abstract}
	This work studies how a classical system set up according to rules informed by a special graph class so that it spontaneously associates to a state space that mimics that of a quantum system. We refer to the states and classical system as quantum-like (QL). The graph plays a dual role by directing the topology of the classical network and defining a state space that comprises superpositions of states in a tensor product basis. The basis for constructing QL graphs and their properties is reviewed and extended. An optimization of the graph product is developed to produce a more compact graph with the essential properties required to produce states that mimic many of the properties of quantum states.  This provides a concrete visualization of the correlation structure in a quantum state space.  The question of whether and, if so, how, entanglement can be exhibited by these QL systems is discussed critically and contrasted to the concept of `classical entanglement' in optics.
	
	\textcolor{green}{This manuscript was prepared for a workshop. It is intended to be a collection of notes on the subject of QL states. Note that some of the notes are speculative, some are being explored further. Gradually the main topics summarized here are being published.}
	
	\end{abstract}

	\maketitle
    \newpage

\tableofcontents

\section{Introduction}

In recent work we have described how a classical system can be mapped, via a graph representation, to a state space that has many properties similar to a quantum state space\cite{ScholesQLstates, QLproducts}. For that reason, we call it a quantum-like (QL) state space, and refer to the states as QL states. The work is broadly motivated by the question of whether there exists any classical systems that can mimic the states of composite quantum systems. In the present work, we describe the construction and some properties of QL states in detail. With the technical details established, we discuss the question of whether useful entangled states can, in principle, be produced by these classical constructions. 

Let's start by defining more clearly what we mean by QL and the motivating idea of mimicking the states of composite quantum systems. We identify classical systems that combine by a special product operation to produce new, exponentially larger, classical systems that are characterized by states that have many similar properties to quantum states. We are not taking a classical system and `quantizing it', nor are  exhibiting classical systems, that we claim to represent quantum systems like electrons or photons. Instead, the QL constructions are special classical systems that produce \emph{state spaces} inspired by composite quantum systems such as many-electron atoms or entangled photons. 

How can we accomplish this, since we do not usually associate classical systems with a ladder of states? The way to think about the strategy is that it is mediated by a graph (defined below). The graph has structure and topology which can serve as a blueprint for the classical system. Essentially, we need to define how correlations or couplings in the classical system need to be organized. Any graph, on the other hand, is associated with a matrix (we use the adjacency matrix). Hence, the graph connects the classical system it designs to a linearized state space. One main technical elaboration is required. To ensure the states are provided in the tensor product basis, we build the composite classical systems based on the Cartesian product of graphs. 

Is it feasible that such classical systems exist? The following argument suggests the answer is yes, and in fact, hints that there might even be better techniques for finding them than described in the present work. Consider the states of electron spin---spin eigenfunctions---in an atom or molecule. For many electrons, these eigenfunctions tend to be tedious to work out in practice\cite{Pauncz}, but their structure is dictated by the Dirac identity, applied to the set of basis functions (product functions, eigenfunctions of $S_z$). The Dirac identity shows that the simultaneous eigenfunctions of $S^2$ and $S_z$ are dictated by irreducible representations of the symmetric group. There are many classical systems that, similarly, are governed by the structure of the symmetric group and, therefore, whose states (if we can define them) will likely be good mimics of the many-electron spin states.  This approach might be too broad, since every group can be represented by the symmetric group (Cayley's theorem), and we might want to tighten the constraints. But nevertheless, it shows feasibility for the search for QL systems in the classical world. 

The graphs needed as a basis for the maps are called QL bits. Particular constructions of these special graphs produce states that, remarkably, have the properties of quantum states. That is, the states for a system of QL bits comprise (controllable) superpositions of basis states, and the basis is a tensor product basis. In a nutshell, we have exhibited a graph with an emergent state vector that transforms as a representation of $\mathsf{SU}(2)$. Then we produce graphs using a Cartesian product operation so that the emergent state vectors transform as $\mathsf{SU}(2) \otimes \mathsf{SU}(2) \otimes \dots$. In the present paper we summarize and extend the background material and survey properties of these states. The question of whether and, if so, how, entanglement can be exhibited by these QL systems is critically discussed. 

The QL bit graphs are not simply a pair of coupled oscillators, but are more sophisticated so that the states are robust\cite{ScholesEntropy} and the spectrum is controlled. That is explained in detail in Sec. V. Furthermore, the QL bits are combined by a special product operation to produce the general QL state space. The QL properties are specific to the state space produced by the graph, not the underlying classical system. Importantly, the states generated from QL graph products have the form of superpositions of product states, and might, therefore, allow a classical realization of  nonseparable states. 

\begin{figure}
	\includegraphics[width=10 cm]{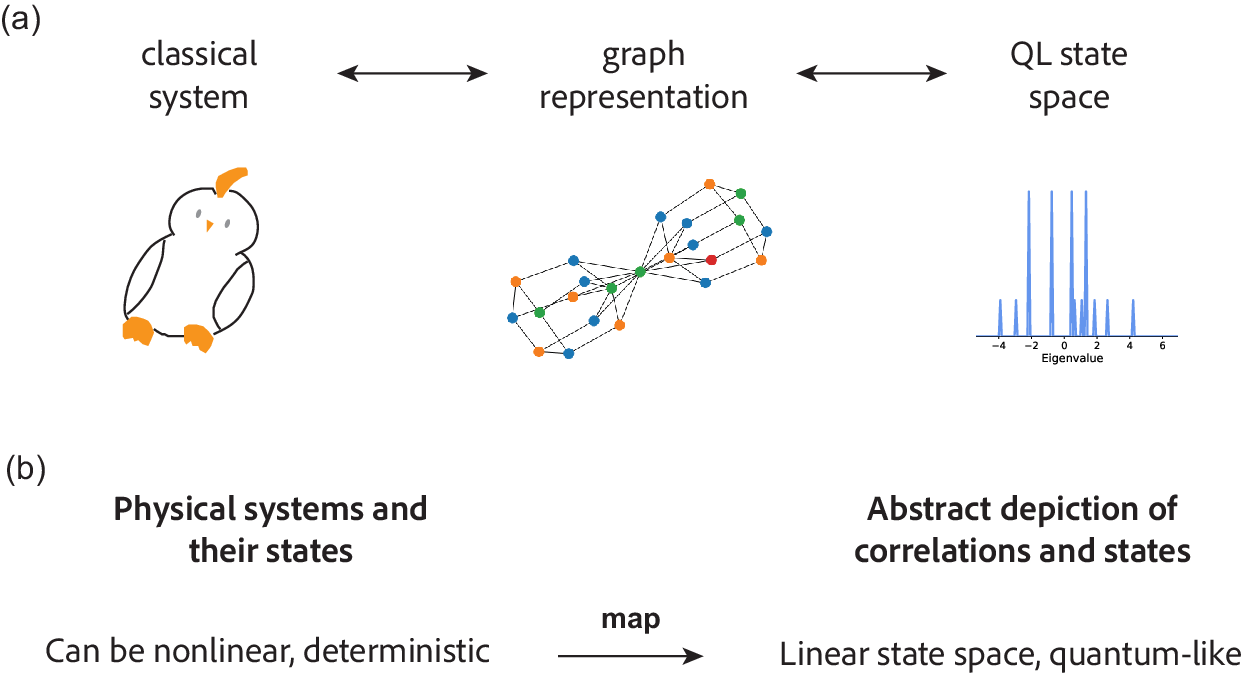}
	\caption{(a) This work studies maps between a suitably designed classical system (here Schrödinger's cat has been reinterpreted as a cockatoo by the artist) and a `quantum-like' (QL) state space that mimics attributes of the state space of a quantum system. The map is mediated by a graph. The graph provides an abstract interpretation of the classical system allowing us to associate a state space with the classical system. (b) This perspective involves a mapping that takes properties of a classical system encoded with a suitable topology of phase relationships to a representation in a state space that has similar properties to a quantum state space.}
	\label{figColloq1}
\end{figure}

The broad idea of the concept we propose is sketched in Fig. \ref{figColloq1}. In Proposition 1 of ref \cite{QLproducts} we established a key result; a map from the Cartesian product of graphs to states with a tensor product basis. The resulting insight enables a graph construction---which could be abstract or could be concretely associated to a classical network---to be mapped to a state space that mimics the state space enjoyed by quantum systems. The result opens up  a way to take maps on the state space (e.g. quantum gates) and map them back to maps on the graph, which, in turn, can represent a physical, classical system, as we recently reported\cite{Amati1}. As we show here, the graph can be reduced to a diagram that is a physical map of the correlations in the QL state space, such as those same correlations that can give rise to entanglement in a quantum state. Here we use this concrete depiction of correlations extensively, aiming to elicit physical insights into how to think about QL states and their properties in the context of the systems that produce those states. On this basis we hope future work can suggest insights into questions concerning interpretation and interplay of quantum mechanical states and systems.

Specific goals of the present paper include:  To summarize the construction and properties of the QL bit graph and how it is used to give insight into quantum states and quantum correlations; To explain how QL bit graphs are combined to produce a product graph, representing states in the product basis, then explore how the graph embeds correlations; To discuss how entangled states can be produced by classical systems and contrast this development to non-separable states of classical light. Ultimately, an open question is how quantum are the QL states? 

\section{Background and Motivation}

In early work on quantum-like systems, Santos showed that a C*-algebra can be defined for a random variable, then the algebra has a representation in a Hilbert space. This allowed development of a formalism for stochastic systems that has a very similar structure to analogous quantum postulates\cite{Santos1974}. Cantoni focused on the states and observables of an arbitrary physical system and studied the generalization of quantum mechanical transition probability\cite{Cantoni1975}. Kaaz compared quantum logic---in terms of the lattice of subspaces of a Hilbert space---to classical probability theory\cite{Kaaz1987}. Using the concept of \emph{fuzzy sets} (sets with a graded continuum membership\cite{Zadeh1965}), a QL uncertainty relation was proposed. Khrennikov initiated an extensive program exploring how and where QL probability laws can be found in diverse fields that include cognition, psychology, and finance\cite{Khrennikov2003, Khrennikov2005, Khrennikov2005b, Khrennikov1, Ozawa2020, KB2013, Khrennikov2016, Khrennikov2018SciRep, Khrennikov2023book}. At the core of that work is the identification of a kind of interference effect that can arise in probabilistic systems because of the way context changes between measurements. Other relevant work includes the cellular automata models studied by Elze\cite{Elze2017, Elze2024}, which can show intriguing correspondences to quantum mechanical systems. 

The present work focuses on the idea that one can associate a state space with many classical systems. In particular, if we can interpret a classical system using a graph, then the state space is naturally defined as the states of the graph\cite{Scholes2020, ScholesQLstates, Excitonics2024}. In this sense the graph provides the mathematical map connecting a classical system to a state space. The inspiration for such an approach comes from representation theory in mathematics\cite{Hall2015, Sagan, Feit}. The idea is that one area can be advanced by finding a similar structure in a quite different area. For example, representation theory exploits the ways that vector spaces can exhibit the same properties as much more abstract algebraic structures. Then we can study the abstract system by using well-known transformations in the corresponding vector space representation. A simple example is to represent the group algebra of $\mathbb{Z}_6$ as an appropriate matrix, such that every element in $\mathbb{Z}_6$ is expressed as a power mod 6 of the matrix. We can do that in many ways. We might choose a two-dimensional square matrix over $\mathbb{R}$ (a representation in $M_2(\mathbb{R})$), generated by 
\begin{equation*}
	\begin{pmatrix}
		\frac{1}{2} & -\frac{\sqrt{3}}{2} \\
		\frac{\sqrt{3}}{2} & \frac{1}{2}  
	\end{pmatrix} .
\end{equation*}
Or we could choose a representation in $M_3(\mathbb{Q})$, 
\begin{equation*}
	\begin{pmatrix}
		0 & -\frac{1}{2} & 0 \\
		0 & 0 & -2 \\
		-1 & 0 & 0
	\end{pmatrix} .
\end{equation*}
A more ambitious example is the Langlands program, where the aim is to use infinite dimensional representations of Lie groups to obtain new insights into number theory\cite{Langlands}. The connection, which has been pursued for decades, requires a complex ``mathematical Rosetta stone'' to enable the translation between disparate fields. Quoting from Ref. \cite{Langlandsbook}: ``Those visionary conjectures have exposed, quite unexpectedly, the deeply entwined nature of several seemingly unrelated branches of mathematics.''

The feature that makes the state space \emph{quantum-like} is a graph construction that allows us to define building blocks that are effectively two-state systems (or, more generally, $n$-state systems). We call these graphs QL bit graphs\cite{ScholesQLstates}. These graphs enable us to represent the properties of a QL state space by a classical system that can be arbitrarily complex.

The QL states we develop can be grouped  in a class of classical systems with potential to display non-classical properties,  together with the classical nonseparable states of light\cite{Spreeuw1998, nonseplight, Forbes2015}. Those states of classical light are local, which is considered a key distinguishing feature for quantum versus classical non-separable  states of light. Although classical non-separable states can show a violation of Bell's inequality, this correlation is thought to lose its significance if the entangled systems cannot be separated\cite{Spreeuw1998, Paneru2020, HarriganSpekkens, nonlocal1, nonlocal2, nonlocal3, nonlocal4, nonlocal5, nonlocal6}. We discuss these issues later in the paper when we analyze entanglement in QL states.

\section{The properties of graphs that generate quantum-like states}

Quantum-like (QL) states are generated by extensive coupling among entities in a network. The idea is general in the sense that these coupled entities can be oscillators\cite{Scholes2020, Strogatz2000}, voltage spikes in a neural network\cite{SyncSpiking}, elements of a game linked by rules for play\cite{Pena}, reactive species in kinetic interplay\cite{bacteriallysis}, interacting people in a social network\cite{Estrada}, and so on. In each of these cases we can represent the entities as vertices of a graph and the coupling (or connections, interplay, or interactions) as edges. The graph structure therefore provides a  framework that unites disparate physical models\cite{WattsStrogatz}.

What is a graph? A graph $G(n,m)$, that we often write simply as $G$, comprises $n$ vertices and a set of $m$ edges that connect pairs of vertices. In one specific case, the vertices denote people in a social network and the edges indicate connections between pairs of people who are friends. The size of a graph or subgraph, that is, the number of vertices, is written $|G|$. The spectrum of a graph $G$ is defined as the spectrum (i.e. eigenvalues in the case of a finite graph) of the adjacency matrix $A$ associated with the graph. The spectrum of a graph is obtained by diagonalizing the corresponding adjacency matrix. The adjacency matrix rows and columns are indexed by the graph vertices. The diagonal entries are zero, while off-diagonal entries contain 1 at $a_{ij}$ if the vertex $i$ is linked by an edge to vertex $j$. Here we will specialize to undirected graphs, so the adjacency matrix is symmetric ($a_{ij} = a^*_{ji}$), but it is also possible to work with directed graphs. For background see ref \cite{Excitonics2024}. We can elaborate this basic model and assign a phase (or bias) to the edges so that they may take any value on the unit circle in the complex plane\cite{Zaslavsky1, Mehatari, Reff2012}. We will exploit this freedom in the present work to present a graph where continuous edge bias rotations over an edge topology transform the emergent state vector according to $\mathsf{SU}(2)$ group operations. For background on graph theory see \cite{Diestel,Janson2000, Bollobas2001}. 

Our goal is to design graphs that can represent a two-level system and that can be used as a resource to generate an exponentially large state space. In order for the state space to be a useful resource, the spectrum of a graph serving as a building block for our structures should contain a single prominent \emph{emergent} state. The graph contains many vertices, therefore the spectrum contains many eigenvalues. It is important that a single `privileged' eigenvalue is clearly distinguished in the spectrum, and therefore separated in eigenvalue from all the other states. This is why the state should be an emergent state. We achieve that using an expander graph. See refs \cite{expandersguide, Expanders, Alon1986, Lubotzky, Excitonics2024} for a starting point for definitions, background, and properties of expander graphs. 

In Sec. V, the definition and some properties of expander graphs are explained in more detail for those readers who are interested. In particular, we address why expanders are powerful for producing QL states. In other words, why we do not simply propose a coupled oscillator model? We have tried to present the material using a mathematical focus, but with explanation that highlights the relevant physics. There are a couple of preliminary points to note. First, for the applications to finite-sized QL states we do not really need the full power of expander families; we simply need graphs with a reliable and sizable spectral gap between the emergent state and all other states. Moreover, we are working with small graphs, so the asymptotics are less directly relevant. However, it is likely that the properties of expander graphs will be useful in future work to enable proofs of theorems concerning the physics of QL states, or, in particular, if we wish to take continuum limits for the graph vertices. Second, in the literature and in this paper, the emergent state usually has the highest eigenvalue. Equivalently it could be lowest eigenvalue; simply reverse the signs of the off-diagonal entries in the adjacency matrix.

In the following discussion $d$-regular graphs are important:
\begin{Definition}
	($d$-regular graph) A graph $G$ is $d$-regular if every vertex has degree $d$. That is, every vertex connects to $d$ edges.
\end{Definition}

A $d$-regular graph is not necessarily an expander graph, but we often use the term to mean $d$-regular \emph{random} graph, which is likely to be an expander (see Sec. V).

\section{The quantum-like bit}

Using the expander graph construction for a network, we have a source of classical states. Now we need to exhibit a graph that can act as a two-level system, and moreover, where we can put the two levels into superposition. In prior work\cite{ScholesQLstates}, we showed a way to accomplish this by coupling two expander graphs into a new graph, the QL bit. One of the expander subgraphs, $G_{a1}$, in the graph $G_A$ can represent one of the states of the QL bit, while the other graph, $G_{a2}$, represents the other state. Alternatively, we can choose a different basis, so that the two states are superpositions of the subgraph states\cite{Amati1}.

\begin{figure}
	\includegraphics[width=6.5 cm]{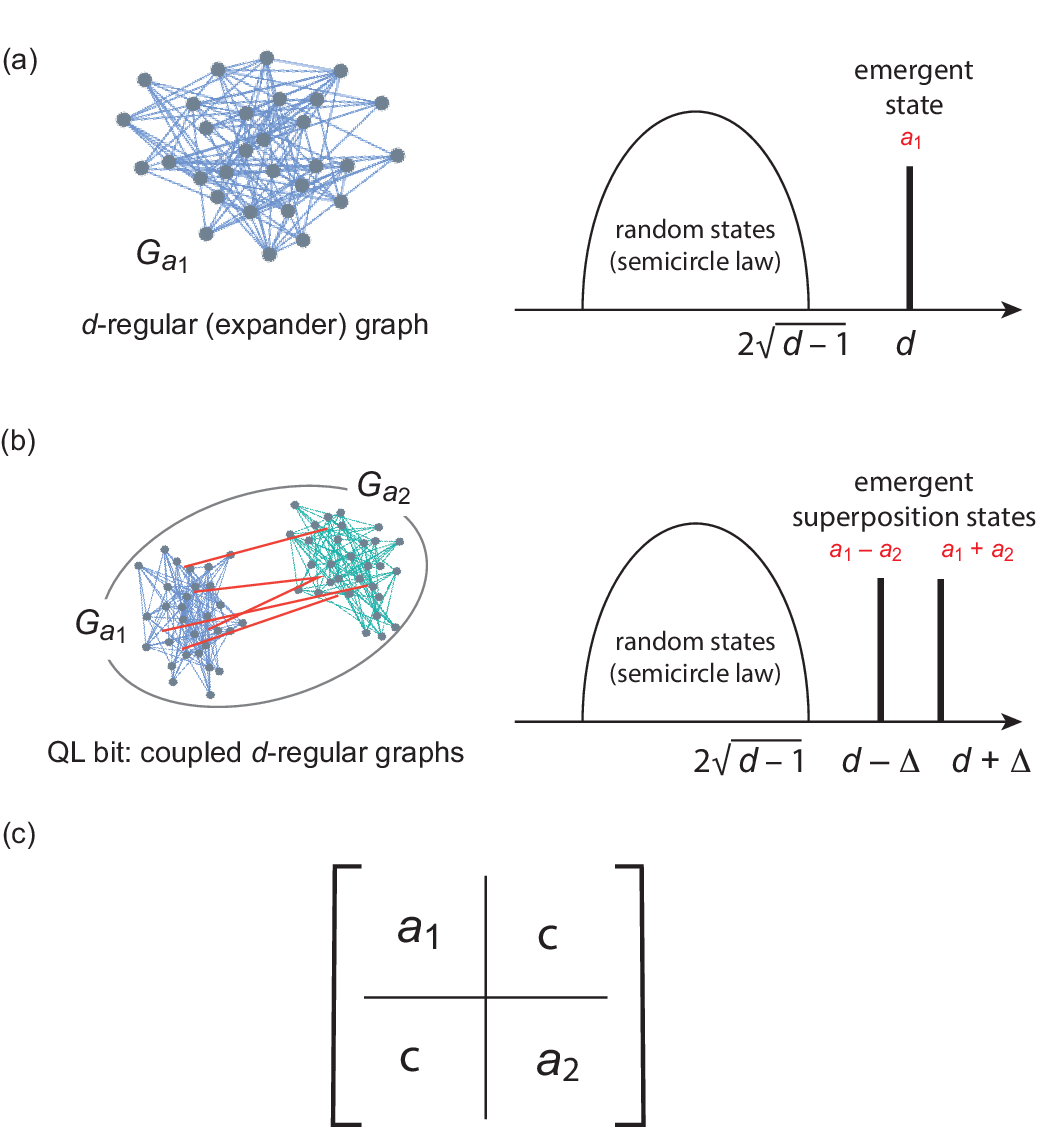}
	\caption{(a) Drawing of a small $d$-regular graph and the spectrum representative of a large $d$-regular graph, showing that the single emergent state is separated in the spectrum from the many other states that we refer to as `random states'. (b) A QL bit is constructed by coupling together two $d$-regular subgraphs. The coupling edges, shown in red, are added randomly from each vertex in $G_{a1}$ to each vertex in $G_{a2}$ with probability 0.2. Realistically, the QL bit will likely not show the subgraphs separated in space, like we display here for clarity; instead the vertices can be positioned randomly. (c) Adjacency matrix of a QL bit showing the diagonal blocks hosting the adjacency matrices for each subgraph. These blocks are coupled by edges in the off-diagonal blocks labeled $c$ that hybridize the subgraphs.}
	\label{figColloq2}
\end{figure}

Specifically, to produce a QL bit, we combine two expander graphs by coupling them using connecting edges, Fig. \ref{figColloq2}. As long as we include a small number of coupling edges compared to the edge density in each of these subgraphs, we can thereby hybridize the emergent states, producing two emergent states that are the in- and out-of-phase linear combinations of the emergent states for each subgraph in isolation. In our examples, we usually connect two $d$-regular (random) subgraphs by 20\% of the edges that would be needed to produce one large $d$-regular graph on all the vertices combined.

The eigenstates of each isolated subgraph are defined in the basis of the $n$ vertices of the relevant graph. For example, for subgraph $G_{a1}$ the emergent state is $|a_1\rangle = (u_1, u_2, \dots, u_n)/\sqrt{n}$, where the $u_i$ are basis states indexed by the graph vertices. Setting each subgraph to contain $n$ vertices, the $2n$-dimensional basis for the QL bit is the tensor sum of the bases of the subgraphs, the basis of $G_{a1}$ and that of $G_{a2}$. In this QL bit basis, then when $d$ is sufficiently large, the hybrid states are approximately $A_- = (|a_1\rangle - |a_2\rangle )/\sqrt{2}$ and $A_+ = (|a_1\rangle + |a_2\rangle )/\sqrt{2}$. If the connecting edges are signed positive, then $A_+$ has the highest eigenvalue. Conversely, if the connecting edges are signed negative, then $A_-$ has the highest eigenvalue. 

Real-valued eigenvector coefficients for the QL bit with $2n$ vertices are those associated to the standard adjacency matrix of the graph $G_A$, that is, where each undirected edge is indicated as 1 in the adjacency matrix $A_G$. We diagonalize $A_G$ and the states associated with the two largest eigenvalues are the emergent states.   Below we provide more detail about how to project these states to states of an effective two-level system. Note that the edges do not have to correspond to a value of 1 in $A_G$. One obvious alternative is to set edge values to $-1$, which means physically that the vertices connected by that edge are perfectly `out-of-phase'. We can also choose any complex number\cite{QLsync}. We could even choose a function\cite{Isospectral}.

The QL bit described and referred to in the remainder of the paper is just one possible design. For example, another way of producing these states we call the type-2 QL bit. Here we construct the QL bit by 2-lifting a $d$-regular graph to produce a bipartite graph\cite{Linial2002, Linial2006, Spielman2015}. We can perform the lift, or simply take $n$ unconnected vertices for $G_{a1}$ and $n$ unconnected vertices for $G_{a2}$. Then couple these vertex sets with many edges so that we produce a bipartite $d$-regular graph. Now our emergent states are the largest and smallest eigenvalues. Furthermore, they should be perfect superpositions when projected into the effective basis, provided both subgraphs have the same number of vertices.

Our basis for the states of the QL bit is defined by the vertices of the graph, $G_A$, which we enumerate in terms of each subgraph as basis states $\{ u_1, u_2, \dots, u_n\}$ for subgraph $G_{a1}$, and $\{x_1, x_2, \dots, x_k \}$ for subgraph $G_{a2}$, given $|G_{a1}| = n$ and $|G_{a2}| = k$. Taken together, we have a complete orthonormal basis in $\mathbbm{R}^{(n+k)}$ or, more generally, $\mathbbm{C}^{(n+k)}$.

An arbitrary emergent state, $W$, of the QL bit graph will be written in terms of this basis as:
\begin{equation}
	W = c_1u_1 + \dots + c_nu_n + d_1x_1 + \dots + d_kx_k
\end{equation}
where the $c_i$ and $d_j$ are complex coefficients. Now define

\begin{eqnarray}
	J_{a1} = \{ \underbrace{1, 1, \dots, 1,}_{\text{\emph{n} times}} \underbrace{0, 0, \dots, 0}_{\text{\emph{k} times}}  \}/\sqrt{n} \\
	J_{a2} = \{ \underbrace{0, 0, \dots, 0,}_{\text{\emph{n} times}} \underbrace{1, 1, \dots, 1}_{\text{\emph{k} times}}  \}/\sqrt{k}.
\end{eqnarray}
We resolve the coefficients for the effective two-states of the QL bit in terms of the basis associated to the subgraphs $G_{a1}$ and $G_{a2}$, $|a_1 \rangle$ and $|a_2 \rangle$, using the inner products
\begin{eqnarray}
	\alpha = \langle J_{a1}, W \rangle \\
	\beta = \langle J_{a2}, W \rangle 
\end{eqnarray}
to give
\begin{equation}
	W_{2 \times 2} = \alpha |a_1 \rangle + \beta |a_2 \rangle .
\end{equation}

We can also design QL graphs that produce single emergent states. These graphs are close mimics of systems like electrons. The nature of vertices might be abstract, but they could physically represent, for instance, circular phase oscillators (grouped as clockwise and anti-clockwise), current oscillations, polarizations, field oscillations or flows, and so on.

To accomplish this, we make the \emph{entire} QL bit graph $d$-regular, but collect the vertices into two groups. For reference, we might call these the red vertices and the blue vertices. We do not need to be able to differentiate vertices among the groups (i.e. red from blue), but we need to specify edge biases within each group and between the groups, up to an overall arbitrary phase. With this edge topology on the graph, we can continuously rotate the edge biases in terms of two phase angles so that the states $W_{2 \times 2}$ sweep through all possible projections on the Bloch sphere. In other words, the graph represents the continuous group operations of $\mathsf{SU}(2)$ on the states. The same is true of the QL bit described above when we focus just on the emergent state with greatest eigenvalue magnitude. 

An example of three orthogonal projections is summarized in Table 1. Here red (blue) bias means the bias on the edges connecting red (blue) vertices, whereas the connecting bias is the bias on edges that connect red to blue vertices. The eigenvalue is indicated as $d$ (or $-d$) because the overall graph in each case is $d$-regular. The red vertices are associated with subgraph $a_2$ and the blue vertices with $a_1$. To obtain uniform eigenvalues of $|d|$ during the continuous edge transformations we need to take some care.  Overall, the graph is $d$-regular random, which means that a quarter of all edges connect the red vertices, a quarter of all edges connect the blue vertices, and half the edges connect these two subgraphs to each other. To retain a $d$-regular graph in the $z$-projection, rather than simply deleting these connecting edges we instead transfer them into the remaining connected subgraph. For instance, for the projection that gives the blue vertex basis state $|a_2\rangle$,  each pair of edges connecting blue to red vertices is pulled back to connect only the two blue vertices in this pair. That results in a $d$-regular subgraph.

\begin{table}[b]
	\caption{\label{tab:table1}%
		An example of a set of QL graph edge biases corresponding to three orthogonal representations of the QL state for a QL bit with a single emergent state.
	}
	\begin{ruledtabular}
		\begin{tabular}{cccccc}
			\textrm{Projection}&
			\textrm{Red bias}&
			\textrm{Blue bias}&
			\textrm{Connecting bias}&
			\textrm{Eigenvalue}&
			\textrm{State}\\
			\colrule
			$x$ & +1 & +1 & +1 & $d$ & $|a_2\rangle + |a_1\rangle $ \\
			$x$ & -1 & -1 & +1 & $-d$ & $|a_2\rangle - |a_1\rangle $ \\
			$y$ & +1 & +1 & $i$ & $d$ & $|a_2\rangle + i|a_1\rangle $ \\
			$y$ & -1 & -1 & $i$ & $-d$ & $|a_2\rangle - i|a_1\rangle $ \\
			$z$ & 0 & +1 & 0 & $d$ & $|a_2\rangle $ \\
			$z$ & -1 & 0 & 0 & $-d$ & $|a_1\rangle $ \\
		\end{tabular}
	\end{ruledtabular}
\end{table}

\section{Graph products that generate quantum-like states}

The next step is that we need to generate superpositions of states in a product basis. We thus need to work out how to combine the QL bit graphs so that a basis for eigenstates of a multi-QL bit system is a tensor product basis. In other words, the state space for each QL bit $i$ is defined in the Hilbert space $\mathcal{H}_i$, where the states are vectors such as $W_{2 \times 2} = \alpha|a_1\rangle + \beta|a_2\rangle$. A state space for $q$ QL bits is
\begin{equation}
	\mathcal{H} = \mathcal{H}_1 \otimes \mathcal{H}_2 \otimes \dots \otimes \mathcal{H}_q
\end{equation}

Our basis is then the set of $2^q$ states
\begin{equation}
	|a_i\rangle \otimes |b_j\rangle \otimes |c_k\rangle \otimes \dots ,
\end{equation}
where $i, j, k, \dots \in \{1, 2\}$ and states $a_i$ come from subgraphs of graph $G_A$, states $b_j$ come from subgraphs of graph $G_B$, and so on.

In prior work\cite{QLproducts} we showed how to construct a one-to-one map between this tensor product basis of states of two-level systems and the states generated by an operation on the QL bit graphs called the Cartesian product\cite{GraphProducts}. We define and explain the graph Cartesian product in the next subsection. The key result is that, given an eigenvector, say $v_A$, for the emergent state of one QL bit with graph $G_A$ and an eigenvector $v_B$ for the emergent state of another QL bit with graph $G_B$, then we have an eigenvector $v_A \otimes v_B$ for the graph product $G_A \Box G_B$. Hence we have a map between a graph---the product graph---and a tensor product in the state space. This product operation on the graph enables the mappings shown in Fig. \ref{figColloq1}.

\subsection{Graph products}

We can produce new graphs from existing graphs using a product operation\cite{GraphProducts}. This allows systematic propagation of a property possessed by the base graphs. For example, in the case of the Cartesian product of graphs $G$ and $H$, described below, the chromatic number of the product graph is that of $H$ or $G$ (whichever is the larger chromatic number). For the present work, we want to propagate the way the QL bit graph represents a two-state system. Graph products are defined starting with the vertex set of the Cartesian product of vertex sets,
\begin{equation*}
	V(G) \times V(H) \rightarrow \underbrace{ \{(u,x) \} }_{\text{set of ordered pairs}},
\end{equation*}
where the set of ordered pairs enumerates all pairs of vertices, one taken from $V(G)$ and one from $V(H)$. Edges are determined by rules governed by the type of product\cite{GraphProducts}. There are four common graph products: the direct product $\times$ (also called tensor product or Kronecker product), the Cartesian product $\Box$, the strong product $\boxtimes$, and the lexicographic product $G[H]$.  The Cartesian product of graphs is the product relevant for our purposes. It is defined here:

\begin{Definition}
	(Cartesian product of graphs) $G \Box H$ is defined on the Cartesian product of vertex sets, $V(G) \times V(H)$. Let $\{u, v, \dots\} \in V(G)$ and $\{x, y \dots\} \in V(H)$. Let $E(G)$ and $E(H)$ be the set of edges in $G$ and $H$ respectively. The edge set of the product graph $G \Box H$ is defined with respect to all edges in $G$ and all edges in $H$  as follows. We have an edge in $G \Box H$ from vertex $(u,x)$ to vertex $(v,y)$ when
	\begin{itemize}
		\item either there is an edge from $u$ to $v$ in $G$ and  $x = y$,
		\item or there is an edge from $x$ to $y$ in $H$ and  $u = v$.
	\end{itemize}
\end{Definition}

The  spectrum of the Cartesian product (see, for instance \cite{Barik2018}) is given by

\begin{Proposition}\label{eq:eig_prod}
	(Spectrum of a Cartesian product of graphs) Given
	\begin{enumerate}
		\item[] A graph $G$, for which its adjacency matrix $A_G$ has eigenvalues $\lambda_i$ and eigenvectors $X_i$, and
		\item[] A graph $H$, for which its adjacency matrix $A_H$ has eigenvalues $\mu_i$ and eigenvectors $Y_i$, then
	\end{enumerate}
	the spectrum of $G \Box H$ contains eigenvalues $\lambda_i + \mu_j$ and the corresponding eigenvectors are $X_i \otimes Y_j$.
\end{Proposition}

\begin{figure*}
	\includegraphics[width=13.5 cm]{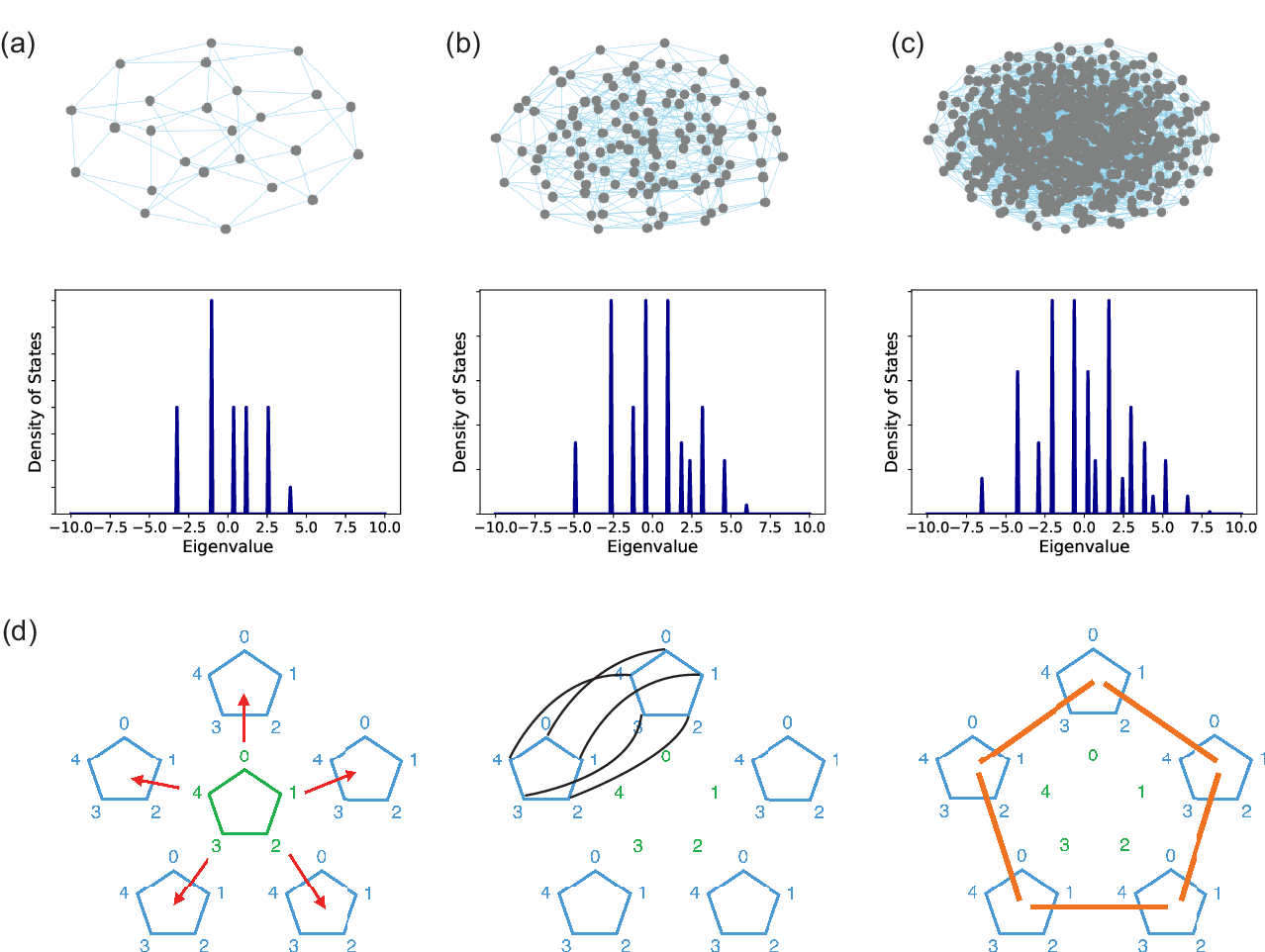}
	\caption{Examples of graph Cartesian products and corresponding spectra for (a) $C_5 \Box C_5$ (b) $C_5 \Box C_5 \Box C_5$ (c) $C_5 \Box C_5 \Box C_5 \Box C_5$. (d) Procedure for the physical construction of  the product $C_5 \Box C_5$, see text. The vertices of each base graph ($C_5$) are labeled $0, 1, \dots, 4$. The vertices of the product graph $G \Box H$ are, accordingly, $(i,j)$, where $i$ labels a vertex in $G$ and $j$ labels a vertex in $H$. }
	\label{figSync3}
\end{figure*}

Let's consider an example of spectra of graph products based on a model cycle graph on five vertices ($C_5$). Spectra of products of the 5-cycle, $C_5$, are displayed in Fig. \ref{figSync3}a-c. The largest eigenvalue of $C_5$ is $\lambda_0(C_5) = 2$, and the second eigenvalue $\lambda_1(C_5) = 0.62$. Thus, for the products we find $\lambda_0(C_5 \Box C_5) = \lambda_0(C_5) + \lambda_0(C_5) = 4$ and $\lambda_1(C_5 \Box C_5) = \lambda_0(C_5) + \lambda_1(C_5) = 2.62$, and so on. Notice that the gap between the highest two eigenvalues, $\lambda_0 - \lambda_1$, remains constant as we take products, so that when we take products of expander graphs, emergent states in the base graphs are emergent states in the product. 

In Fig. \ref{figSync3}d we show how the product $C_5 \Box C_5 = G \Box H$ is produced explicitly. Let's label the graphs $G$ and $H$. One of the graphs, say $H$ (drawn in green), templates the product. For each vertex in the graph $H$ we draw one copy of the the graph $G$ (the blue graphs). The vertices are indexed by the index pair, one corresponding to the graph $G$ and one index deriving from the vertex of $H$ associated with each copy of $G$. We then use these second indices to draw edges between identical vertices of the copies of $G$  templated according to the edges in $H$. One set are shown as the black edges. Finally we have the product graph, which can be drawn to display explicitly the graphs $G$, as shown in the figure, or we could draw a similar picture that clearly shows five copies of $H$. The graph product thereby gives a physical picture of the tensor product.

Notice that the Cartesian graph product $G_A \Box G_B$ is constructed by putting a copy of $G_A$ at each vertex of $G_B$, then completing the additional edges. This mirrors the way a tensor product of $n$-dimensional Hilbert spaces $\mathcal{H}_A \otimes \mathcal{H}_B$ can be viewed as an $n$-fold direct sum of $\mathcal{H}_A$. See Remark 2.6.8 in ref \cite{KadRing1} for a proof. Also note that the method for drawing the product graph indicates how the corresponding adjacency matrix is structured in block form.

It is also evident that the group operations underlying each graph are preserved in the product. To illustrate, consider a product of two different $d$-regular graphs, $X_1$ and $X_2$. We can construct each $d$-regular graph as a Cayley graph\cite{expandersguide}, so that $\text{Cay}(\Gamma, S)$ produces the edges controlled by the generating set $S$ and the graph is $|S|$-regular, where $|S|$ means the size of the set $S$. The vertices of the graph are given by the elements of $\Gamma$ and the edges are defined by the group operation of elements $S \subset \Gamma$, where if $s \in S$ then $s^{-1} \in S$. The group operations for one graph, say $X_1$ are preserved in each subgraph of the product, while the group operations for the other graph ($X_2$) are encoded in the edges connecting the copies of $X_1$. Prior work has  developed the theory for tensor products of groups\cite{BrownLoday, tensorgroup2} and related topics\cite{Ellis1991}. Similarly, we can focus on the way the edge topology of a product graph is generated by the product of edge topologies of each base graph\cite{Gross-Tucker}. This is similar to product space topology more generally\cite{Willard}.

It is instructive to calculate examples of $d$-regular random graphs, which are the basis for the QL bit subgraphs. A $d$-regular random graph on $n$ vertices has a total of $dn/2$ edges, arranged such that each vertex connects to precisely $d$ edges. We produce these graphs as described previously\cite{ScholesEntropy}. In these calculations $d = 8$ and the graph $G(n,m)$ has $n=12$ vertices and $m=44$ edges. Each graph $G(n,nd/2)$ is randomly generated and $nd/2-m=4$ edges are randomly deleted. We do not know precisely how the vertices are connected in each graph, highlighting that the building blocks for QL bits are weakly dependent on the precise details of the subgraphs.  To further emphasize the resilience of these states to disorder\cite{Scholes2020}, we introduce `frequency disorder' by adding values to the diagonal elements of the adjacency matrix from a normal distribution. This is known as a `random Schrödinger operator'\cite{Geisinger}, studied in the context of graph delocalization and quantum ergodicity\cite{Anantharaman1, Anantharaman2}. The spectrum plotted has a standard deviation of the distribution of $\sigma = 2.0$. Compare this value to the off-diagonal entries in the adjacency matrix (`couplings'), which are set to 1.

\begin{figure}
	\includegraphics[width=13.5 cm]{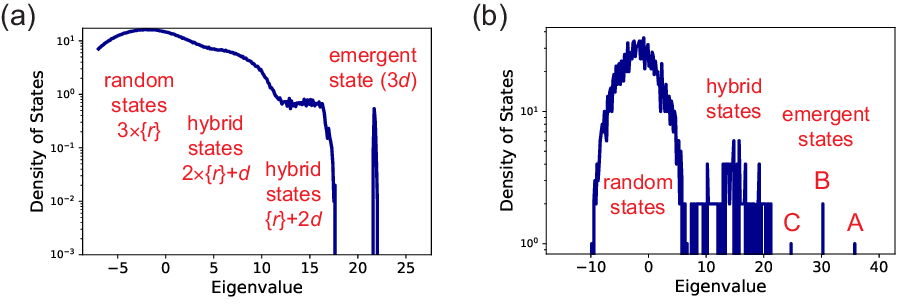}
	\caption{(a) Calculated spectra of $G_1 \Box G_2 \Box G_3$ where the base graphs are disordered $d$-regular random graphs. The plot shows the ensemble spectrum for a graph with diagonal (`frequency') disorder $\sigma = 2.0$. The emergent state is produced by the tensor product of emergent states of each graph. Other states are various products of sets of random states, denoted $\{r\}$, and emergent states, as indicated. (b) Spectrum of the product of two QL bits. Reprinted from G. D. Scholes and G. Amati, 2025, Quantumlike Product States Constructed from Classical Networks, Phys. Rev. Lett. 134:060202.}
	 \label{figSync4}
\end{figure}   

In Fig. \ref{figSync4}a we show an ensemble spectrum of a graph product of these $d$-regular random graphs, $G_1 \Box G_2 \Box G_3$. The three graphs $G_1, G_2, G_3$ differ by the random deletion of the four edges and we take an ensemble average over realizations of these graphs. We identify the set of random states, see labels in Fig. \ref{figSync4}a, which are complemented by bands of `hybrid states' that comprise eigenvalue sums of random-states and one or two emergent-state eigenvalues. The emergent state is prominent, and in the product graph it remains separated from the second eigenvalue (a hybrid state). In Fig. \ref{figSync4}b we show an example of the spectrum of a single product of two QL bits\cite{QLproducts}. Notice the four emergent states, labeled A, B (doubly degenerate), and C. 

\subsection{The effective QL states}

The effective (pure) states obtained for the graph product of two QL bits are written as
\begin{equation}
	W^i_{4 \times 4} = |\Psi\rangle = c^i_1|a_1\rangle |b_1\rangle + c^i_2|a_2\rangle |b_1\rangle + c^i_3 |a_1\rangle  |b_2\rangle + c^i_4 |a_2\rangle  |b_2\rangle ,
\end{equation}
which define any state of $\mathcal{H} = \mathcal{H}_A \otimes \mathcal{H}_B$. The $c^i_k$ are complex coefficients obtained from the relevant projections described below. Given particular states of each QL bit A and B:
\begin{eqnarray}
	|\psi_A\rangle = W^A_{2 \times 2} = \alpha_1 |a_1\rangle + \alpha_2 |a_2\rangle \\
	|\psi_B\rangle = W^B_{2 \times 2} = \beta_1 |b_1\rangle + \beta_2 |b_2\rangle ,
\end{eqnarray}
The graph product yields the separable state $ |\Psi\rangle = |\psi_A\rangle \otimes |\psi_B\rangle$ with $c_1 =\alpha_1\beta_1$, $c_2 =\alpha_2\beta_1$, $c_3 =\alpha_1\beta_2$, and $c_4 =\alpha_2\beta_2$. For elementary background in the context of quantum states see Ref. \cite{BellTutorial}.

We obtain the coefficients as follows. We form the Cartesian product of two QL bit graphs, say $G_A$ and $G_B$, each comprising two subgraphs with $n_0$ vertices. From the product, $G_A \Box G_B$, we end up with four subgraphs (each with $n_0^2$ vertices). The subgraphs are connected with a certain structure of edges that define the correlations. The vertices in each subgraph are labeled by all possible $n_0^2$ pairs of vertex indices, where one index is taken from a subgraph of $G_A$ and the other from a subgraph of $G_B$. 

We project the effective emergent states from the eigenvectors $W_i$ of the entire graph to effective states in a chosen basis. Here we choose the natural product basis associated directly with the subgraphs of $G_A \Box G_B$: $|a_1\rangle |b_1\rangle$, $|a_2\rangle |b_1\rangle$, $|a_1\rangle |b_2\rangle$, $|a_2\rangle |b_2\rangle$, although we are free to choose any basis. The natural basis is selected by defining the vectors
\begin{eqnarray}
	J_{a1b1} = \{ \underbrace{1, 1, \dots, 1}_{n_0^2 \text{ times}} \underbrace{0, 0, \dots, 0}_{3n_0^2 \text{ times}}  \}/\sqrt{N} \\
	J_{a2b1} = \{ \underbrace{0, 0, \dots, 0}_{n_0^2 \text{ times}} \underbrace{1, 1, \dots, 1}_{n_0^2 \text{ times}} \underbrace{0, 0, \dots, 0}_{2n_0^2 \text{ times}}  \}/\sqrt{N} \\
	J_{a1b2} = \{ \underbrace{0, 0, \dots, 0}_{2n_0^2 \text{ times}} \underbrace{1, 1, \dots, 1}_{n_0^2 \text{ times}} \underbrace{0, 0, \dots, 0}_{n_0^2 \text{ times}}  \}/\sqrt{N} \\
	J_{a2b2} = \{ \underbrace{0, 0, \dots, 0}_{3n_0^2 \text{ times}} \underbrace{1, 1, \dots, 1}_{n_0^2 \text{ times}}  \}/\sqrt{N}
\end{eqnarray}
where it is assumed for simplicity that the ordering of the vertices in the graph product partitions the subgraphs in sequence (which may not be the case in numerical work because it depends on how the adjacency matrix is indexed). Then we obtain coefficients for the effective four states of the QL bit product in terms of the product basis  arising from the subgraphs:
\begin{eqnarray}
	c^i_1 = \langle J_{a1b1}, W_i \rangle \\
	c^i_2 = \langle J_{a2b1}, W_i \rangle \\
	c^i_3 = \langle J_{a1b2}, W_i \rangle \\
	c^i_4 = \langle J_{a2b2}, W_i \rangle 
\end{eqnarray}
where $i$ labels the state of the QL bit, ordered by eigenvalue and $\langle \dots \rangle$ means the inner product.

Linear combinations of product states are produced when the basis graphs, for example $G_{a1}$ and $G_{a2}$, are coupled to produce the superposition states described above, $G_{a1 + a2}$, etc., giving the following states (where,  for clarity, the coefficients $c^i_k$ are not written):
\begin{subequations}
    \begin{align}
	    v_{++} =   \quad |a_1\rangle |b_1\rangle + |a_1\rangle |b_2\rangle 
	    + |a_2\rangle |b_1\rangle +  |a_2\rangle |b_2\rangle  \\
	    v_{-+} =   \quad |a_1\rangle |b_1\rangle + |a_1\rangle |b_2\rangle 
	    - |a_2\rangle |b_1\rangle -  |a_2\rangle |b_2\rangle    \\
	    v_{+-} = \quad |a_1\rangle |b_1\rangle - |a_1\rangle |b_2\rangle 
	    + |a_2\rangle |b_1\rangle -  |a_2\rangle |b_2\rangle    \\
	    v_{--} =   \quad |a_1\rangle |b_1\rangle - |a_1\rangle |b_2\rangle 
	    - |a_2\rangle |b_1\rangle +  |a_2\rangle |b_2\rangle .
    \end{align} 
\end{subequations}

This set of states is documented here so we can refer to it later in this paper. The ordering of the states with respect to their eigenvalues, i.e. the index $i$ in $c^i_k$, depends on the edge bias topology of the two QL bits. That is, any of these four states can be the emergent state with highest eigenvalue.

\subsection{Optimizing the product}
The graph Cartesian product increases the vertex set multiplicatively---that is, the product of $q$ QL bits, each comprising $N$ vertices, generates a product graph comprising $N^q$ vertices. Thus, the size of the resource (the graph) scales proportionally with the size of the state space. Since each QL bit might contain many vertices, it is desirable to form the product most economically. A minimal product structure can be produced by contracting the graph as follows.

For simplicity, assume that each QL bit graph is identical. However, the precise structures do not matter in the end because the important correlation structure comes from the edges that connect the subgraphs, not the structure of the subgraphs themselves. As described above, forming the product $G_A \Box G_B$ involves installing a copy of $G_A$ at every vertex of $G_B$, or \emph{vice versa}, then connecting the vertices among $G_A$ graphs as prescribed by the edge structure in $G_B$. 

\begin{figure*}
	\includegraphics[width=13.5 cm]{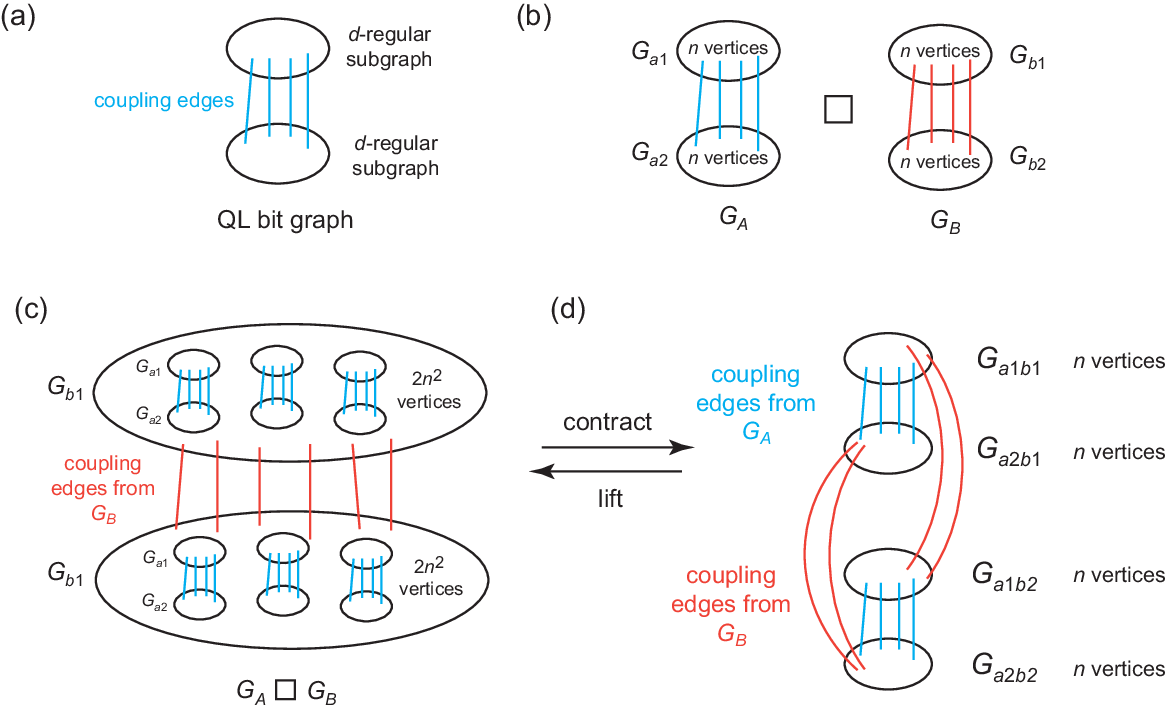}
	\caption{Outline of how the Cartesian product of QL bits, where each $d$-regular subgraph comprises $n$ vertices, produces a much larger graph, where each subgraph comprises $n^2$ vertices and is $2d$-regular. This large graph can be contracted to an optimal graph, where each $d$-regular subgraph comprises $n$ vertices, that retains the qualitative features of the large graph. }
	\label{figSync5}
\end{figure*}

Let's view the product construction for QL bits schematically, Fig. \ref{figSync5}a,b. Recall that each subgraph $G_{a1}$, $G_{a2}$, $G_{b1}$, $G_{b2}$ is $d$-regular. Let's say each subgraph contains $n$ vertices. Notice that after we connect all the copies of $G_A$ that are associated with $G_{b1}$ (or similarly $G_{b2}$), we generate one large $2d$-regular subgraph on $n^2$ vertices by connecting all the $G_{a1}$ subgraphs together. We also produce one large $2d$-regular subgraph on $n^2$ vertices by connecting all the $G_{a2}$ subgraphs. This happens because we connect $n$ subgraphs and each contains $n$ vertices. 

Each vertex inherits the $d$ edges from the original graph (e.g. $G_{a1}$) and attains a further $d$ edges from the edge structure imposed by $G_{b1}$ (or $G_{b2}$), also $d$-regular. This increase in vertex degree is the reason that the eigenvalue of the emergent state of the uncontracted product $G_A \Box G_B$ is found at $2d$ instead of $d$\cite{QLproducts}.

Notice that within the vertex set labeled $G_{b1}$ (or $G_{b2}$) in the product, Fig. \ref{figSync5}c, that we now have a new $G_{a1}$ $2d$-regular subgraph connected by coupling edges to a new $G_{a2}$ $2d$-regular subgraph. We can simplify this graph without changing its properties qualitatively by contracting each $G_{a1}$ and each $G_{a2}$ $2d$-regular subgraph to corresponding $d$-regular sugbraphs. This produces an optimized product, Fig. \ref{figSync5}d. Conversely, the optimized product graph can be lifted to the full product using edge subdivisions. 

A subdivision of an edge connecting vertices $v_i$ and $v_j$ in a graph $G$ produces a new graph by replacing the edge with new connected vertices that connect to each of $v_i$ and $v_j$. These new connected vertices form a subgraph $X$, which might be a single vertex joining $v_i$ and $v_j$, or a more complex connected graph. A contraction of the subgraph $X$ in the graph reverses the subdivision, producing $G$. Similarly, we can apply a contraction to any subgraph in $G$. On the basis of these definitions and the properties of $d$-regular graphs (specifically the isoperimetric property), it is obvious how a $2d$-regular graph on $n^2$ vertices can be contracted to a $d$-regular graph on $n$ vertices. Divide the $n^2$ vertices into $n$ sets each containing $n$ vertices, then contract each of these sets into a single vertex. Concomitantly, the edges between sets collapse so that each contracted set, now a single vertex, has degree $d$. This  yields a $d$-regular graph on $n$ vertices. Conversely, subdivisions of edges of a $d$-regular graph can lift it to a $2d$-regular graph.

Now consider a state comprising $q$ QL-bits. Each QL bit contains $N = 2n$ vertices, $n$ in each subgraph. Therefore, $q$ QL bits contain $qN = 2qn$ vertices whereas their Cartesian product contains exponentially more vertices, $(2n)^q$ vertices in total. There are $2^q$ subgraphs, each comprising $n^q$ vertices. However, after contraction so that each subgraph contains $n$ vertices, we substantially lesson the high resource cost in subgraph vertices so that the overall graph now contains $n2^q$ vertices. 

The reason this dramatic reduction in resource requirement is possible, while still preserving the properties of the product, is that each subgraph is a large $d$-regular graph wherein the precise location of edges and number of vertices are unimportant. All that matters is that the graph is $d$-regular. The important edge construction derives from weakly connecting two $d$-regular subgraphs within each QL bit. From the definition of the Cartesian product of graphs, we have edges between subgraphs labeled $a_i$, $b_j$, $c_k, \dots$ to those labeled $a_l$, $b_m$, $c_n, \dots$ when either $i \neq l$ and $j = m$ and $k = n$, or when $i = l$ and $j \neq m$ and $k = n$ or when $i = l$ and $j = m$ and $k \neq n$. We preserve that important edge topology propagated through the product. Considering these principles, the QL bit graph products have a nice schematic representation, evident in Fig. \ref{figSync5}d.

\subsection{QL graph topology encodes the correlations}

\begin{figure*}
	\includegraphics[width=10 cm]{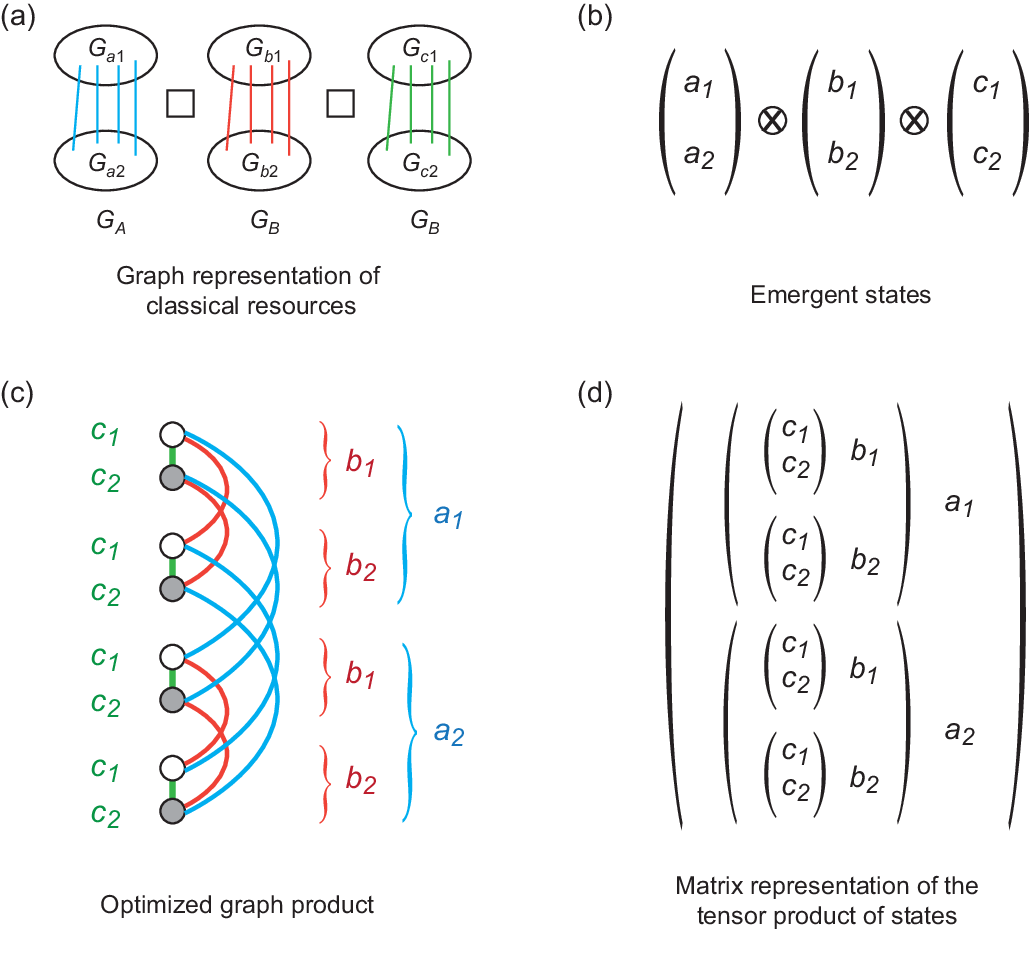}
	\caption{Comparison of (a) the Cartesian product of graphs, $G_A \Box G_B \Box G_C$, and (b) tensor product of states in a three QL-bit system. (c) The schematic representation of the graph product. Emergent eigenstates of associated adjacency matrix comprise the QL state space. (d) Matrix representation of the tensor product frpm part (b). The eigenvectors of this matrix are equivalent to those of the emergent QL states. }
	\label{figSync6}
\end{figure*}

Fig. \ref{figSync6} summarizes the facts we have so far established. The Cartesian product of graphs $G_A$, $G_B$, $G_C \dots$ produces a new graph that has an emergent state that is the tensor product of the emergent states of each of the graphs $G_A$, $G_B$, $G_C \dots$. This key result comes from the definition of the properties of the graph Cartesian product\cite{QLproducts}. The optimized graph product has an edge structure that reveals correlations introduced by superpositions in the tensor product basis. The graph structure is translated to the adjacency matrix, which mirrors the matrix representation of the tensor product of states, Fig. \ref{figSync6}d.

The Cartesian graph product generates a new graph with a special structure encoded in the edge biases. It is this graph topology that generates a QL state space. There are two key ingredients underpinning this outcome. 

First, the subgraphs---that is, the $d$-regular graphs---are the vertices of the correlation structure shown in Fig. \ref{figSync6}c. These \emph{effective} vertices no longer correspond to the original subgraphs, that is, $G_{a1}$, $G_{a2}$, $G_{b1}$, and so on, that they were transcribed from. They now represent the product basis. We can read off those basis states for each subgraph (effective vertex). As shown in Fig. \ref{figSync6}c the basis states from the top down are $c_1b_1a_1$, $c_2b_1a_1$, $c_1b_2a_1$, and so on. The important underlying design principle here is that the constituent graphs in the product ($G_A$, $G_B$, etc.) are two-state graphs. Thus the effective vertices of the optimized product enumerate all permutations of the possible states of our set of QL bits---which is evident in the product basis.  We are free to choose whether the product function is symmetric or antisymmetric under pairwise permutations (see Appendix).

The second key ingredient is the edge structure, or graph topology. That is, the way the subgraphs are connected, shown as the colored edges in Fig. \ref{figSync6}c,  builds the state space by introducing nested correlations between graphs, analogous to the matrix representation of the tensor product, Fig. \ref{figSync6}d. Biasing these connecting edges (that is, allowing edges to take any value on the unit circle in the complex plane) produces a set of linear combinations of the product basis states.

\section{What are expander graphs and why do we use them?}

Expander graphs\cite{Sarnak2004, expandersguide, Lubotzky, Expanders, Expanders2, Alon1986, Tao-expanders} are highly connected graphs that are optimal for random walks and communications networks. The physical concept underpinning an expander is that the edges are scale-free, so that no matter how we lay out the vertices, edges connect vertices at all length scales. A subset of the $d$-regular graphs have this property where edges connect vertices over many distance scales (distance defined with respect to the graph), and are prototypical expander graphs. However, it is not so easy to pinpoint which $d$-regular graphs are expander graphs for reasons that will be clarified below. Nevertheless, $d$-regular \emph{random} graphs are asymptotically almost certainly good expanders\cite{Friedman1991, Miller2008}. These are the graphs we use for the QL bit subgraphs. 

Crucial for the present work is the property that an expander graph has a spectrum with a guaranteed spectral gap between the emergent state and the next eigenvalue. We order the eigenvalues of a graph as $\lambda_0 \ge \lambda_1 \ge \dots \ge \lambda_{n-1}$.  The largest (first) eigenvalue of a $d$-regular graph is $\lambda_0 = d$. It is commonly referred to as the trivial eigenvalue. The second eigenvalue is denoted $\lambda_1$. The gap between $\lambda_0$ and $\lambda_1$  is indicated by the well-known Alon-Boppana bound: Let $G(n,d)$ be an infinite (in the number of vertices, $n$) family of $d$-regular connected graphs on $n$ vertices, with $d$ being fixed. Then, as $n \rightarrow \infty$ the second largest eigenvalue has a bound $\lambda_1[G(n,d)] \ge 2\sqrt{k - 1} - o(1)$. This suggest that there exists a best possible gap, which is a property displayed by expander graphs. 

\begin{figure}
	\includegraphics[width=5 cm]{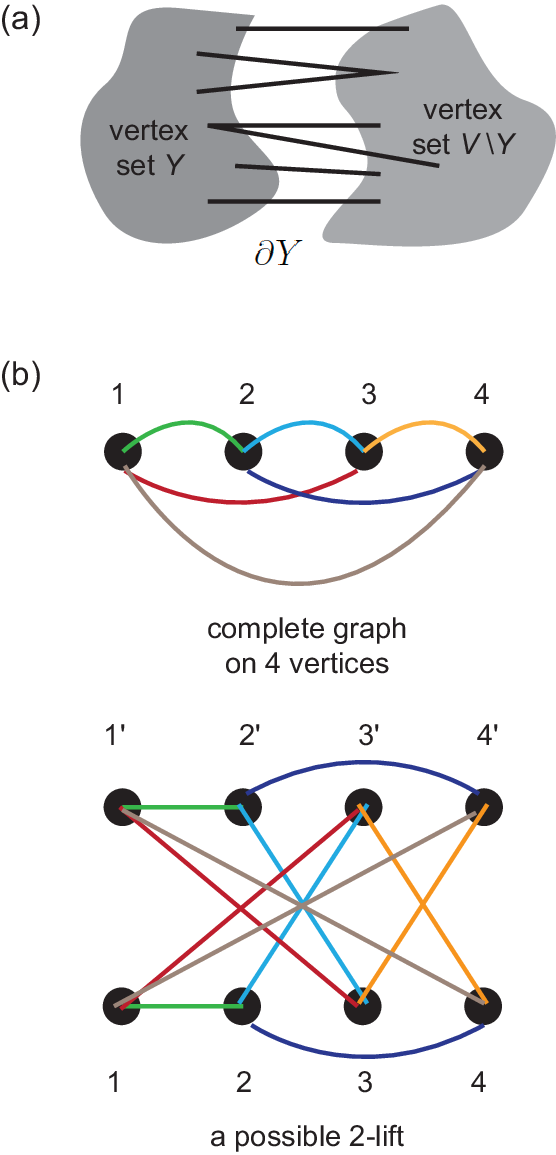}
	\caption{(a) Illustration of a graph divided into two partitions of the vertices. The edges connecting these partitions are the boundary ($\partial Y$). (b) Example of a 2-lift of the complete graph on 4 vertices ($K_4$). Notice that when an edge between vertices $(u, v)$ in the original graph lifts to the pair $(u, v)$ and $(u', v')$, the new graph contains a local connection. Whereas, when it lifts to the pair $(u, v')$ and $(u', v)$, the connection can be long-ranged. }
	\label{figgraph}
\end{figure}   

\subsection{Scales of Edge Connectivity}
How can we understand the manifestation of scale-free edge connectivity in expander graphs? This is an especially interesting concept once you think about the fact the the number of vertices in the graph ($n$) can be very large compared to the number of edges associated with each vertex (i.e. $d$). 

Connectivity throughout the graph is quantified by the isoperimetric constant:
\begin{Definition}
	(Isoperimetric constant) The isoperimetric constant of a graph $G$ with vertex set $V$, is defined as
	\begin{equation}
		h(G) = \min \Big\{ \frac{|\partial Y|}{|Y|}  \Big\} 
	\end{equation}
	with $Y \subset V$ and $|Y| \le \tfrac{1}{2}|V|$ . Here, $\partial Y$ is the boundary of $Y$, which means the set of edges in $G$ that have one endpoint in $Y$ and one endpoint in $V \setminus Y$. 
\end{Definition}

The isoperimetric constant (also known as the Cheeger constant) is a measure of how many bottlenecks a random walk on the graph would encounter. The idea of the isoperimetric constant of a graph is shown in Fig. \ref{figgraph}a. The larger $h(G)$ is, the more connected the graph is. A way to think about it is that, no matter how ingeniously we cut through the graph to divide it into two vertex sets, these vertex sets are always well connected. What is meant by `well connected' is that the number of edges connecting these vertex sets (i.e. $|\partial Y|$) does not diminish faster than $n$ as we look at larger and larger graphs in the family. The concept of isoperimetric constant thus allows us to define an expander graph by considering families of graphs that are highly connected, no matter how large we make them (i.e. how many vertices they contain). These are known as expander graphs. 

Expander graphs are defined in terms of families of $d$-regular graphs with fixed $d$ and any number $n$ of vertices\cite{expandersguide}:
\begin{Definition}
	Let $d$ be a positive integer. Let $(G_n)$ be a sequence of $d$-regular graphs on $n$ vertices such that $|G_n| \rightarrow \infty$ as $n \rightarrow \infty$. We say that $(G_n)$  is an expander family if the sequence $(h(G_n))$ is bounded away from zero.
\end{Definition}

For a family of $d$-regular expander graphs it turns out that a non-zero isoperimetric constant implies a spectral gap. Notably, 
\begin{equation}
	\frac{d - \lambda_1}{2} \le h(G) \le \sqrt{2d(d - \lambda_1)} .
\end{equation}
Or put another way, the persistent spectral gap comes about because the graphs are highly connected across all scales.

The set of all $d$-regular graphs contains prototypical examples of expanders. But not \emph{all} $d$-regular graphs are expanders. For example, any kind of periodic lattice is not an expander. A simple example is a cycle (which is 2-regular). Consider what happens to $h(G)$ as a cycle $G$ becomes infinitely large. Cut the graph into two parts. No matter how you do that, we find $|\partial Y| = 2$ because the graph is a single cycle. Then $h(G) \rightarrow 0$ as the size of the cycle tends to infinity because $|Y| \rightarrow \infty$ as we minimize the ratio $|\partial Y| / |Y|$. The problem with the cycle graph, or any circulant graph\cite{Alon1994, Friedman2006, So2011}, is that, at some $n$ relative to $d$, the edges are essentially local in the sense that they cannot extend throughout the graph because the distances in the graph for large $n$ exceed the period of the edge links. For example, if our graph family starts with $K_8$, then the longest distance that an edge can span is across 8 vertices---which is short once the graph has been expanded to, say, 100 vertices.

Using the principles of Cayley graphs enables us to think about how graphs can be generated following the properties of a group $\Gamma$ (see Sec. 11 of ref. \cite{Expanders} or ref \cite{expandersguide}).This viewpoint, in turn, enables us to link the high connectedness of the graph, expressed by the isoperimetric constant, to the spectral gap, and a special property of the graph eigenvectors\cite{KazhdanTbook, Lubotzky, UnitaryRepbook, Zuk2003}.  We consider how the unitary representations $\pi(s)$ of each group operation $s \in S \subset \Gamma$ act on vectors in a Hilbert space $\xi \in \mathcal{H}_{\pi}$ associated to $\Gamma$. The unitary representation $\pi$ of the group is said to contain `almost invariant' vectors if
\begin{equation}
	\lVert \pi(s) \xi - \xi \rVert < \epsilon,
\end{equation}
for all $s \in S$, where $\lVert \dots \rVert$ means norm (defined by the inner product). This quantifies how much the vectors get `rotated away' from the emergent state vector by the group operations. For expander families, that rotation tends to a non-zero limit as graphs get larger, for every $\pi(s)$. We then say that the graphs have Kazhdan's property (T). Property (T) means that the trivial representation of $\Gamma$, that produces the eigenvalue $d$ for a $d$-regular graph (i.e. the emergent state), is `bounded away' from the other irreducible representations\cite{Lubotzky2005}. That is, the vectors in the Hilbert space are are distinct from the emergent vector under any allowed unitary transformation, so we have a spectral gap. Property (T), like the isoperimetric constant, is hard to determine, so instead we can use the fact of the spectral gap to estimate the constant associated with Kazhdan's property (T)\cite{Zuk2003}.

\subsection{Explicit and Optimal expanders}
As explained above, $d$-regular random graphs are invariably good expanders. However, explicit construction of expander families is  difficult to elucidate, which motivated a study by Bilu and Linial\cite{Linial2006}. By exploiting the technique of random graph coverings, known as random lifts\cite{Linial2002, Linial2006a} they could conjecture how to explicitly construct optimal expander families. The conjecture was later proved for the case of random 2-lifts starting from any $d$-regular complete bipartite graph\cite{Spielman2015}.

The significance of the lifts of graphs is that it suggests a way to generate highly connected graphs. The 2-lift $\tilde{G}$ of a graph $G$ is carried out as follows. Make a copy of the vertices in $G$ and think of this set, labeled $\{u', v', \dots\}$, as lying above the vertices of $G$, labeled  $\{u, v, \dots\}$, . For each edge in $G$ we randomly produce a pair of edges in $\tilde{G}$, chosen as follows. Say there is in edge in $G$ connecting $u$ and $v$. Then the new edges in $\tilde{G}$ can be either the pair $(u, v)$ and $(u', v')$, or the pair $(u, v')$ and $(u', v)$. 

Starting with a complete graph, we can take a succession of 2-lifts to generate an infinite family of optimal expander graphs. How this works is shown in the example of one 2-lift on $K_4$ (the complete graph on four vertices) in Fig.  \ref{figgraph}b. Notice how one kind of edge pair choice introduces local connections, whereas the other manifests global edge connectivity. A sequence of 2-lifts thereby probabilistically builds in edges connecting vertices across the breadth of the graph on all scales; which is precisely what we need. Notably, this can be achieved without needing to crowd the graph with edges. For example, the graph can simply be 4-regular random. It is this interplay between connectivity across all scales of the graph together with sparsity of edges that is the hallmark of optimal expander graphs. 

Proofs are converging on the conclusion that the distribution of non-trivial eigenvalues in the spectrum of $d$-regular random graphs is close to the semicircle law for the eigenvalues of the ensemble of matrices from the Gaussian orthogonal ensemble\cite{Bauerschmidt2017, Sarnak2004, Miller2008}. These are the spectrum of states that we simply refer to, in the context of graphs for QL states, as the \emph{random states}. Studies of $\lambda_1$ suggest it follows the Tracy-Widom distribution\cite{TracyWidom}. Recent work\cite{Huang2022} proved that $\lambda_1/\sqrt{d - 1} \le 2 + \mathcal{O}(n^{-c})$, where $c > 0$ is a constant, which improved the error bound on the Alon-Boppana result. The associated bound on fluctuations of the extreme eigenvalues is further improved in a new paper\cite{Huang2024}, giving $\lambda_1/\sqrt{d - 1} = 2 + n^{-\frac{2}{3} + o(1)}$.  As a consequence of these advances, it is likely that the majority of $d$-regular random graphs are optimal expanders---that is, the spectral gap is close to the best possible. Specifically, it means that 51\% of the bipartite $d$-regular random graphs are optimal expanders, and  27\% of non-bipartite $d$-regular random graphs are optimal expanders.

Expanders can be explicitly constructed, as was first demonstrated by Margulis.\cite{Margulis}. A strategy is to produce a Cayley graph based on a group that has property (T), such as $SL_n(\mathbb{Z})$ with $n \ge 3$ ($SL_n(\mathbb{F}$ denotes the special linear group of $n \times n$ matrices with entries from the field $\mathbb{F}$ and with unit determinant). See, for example, Ref. \cite{Lubotzkybook}. Optimal expander graphs are called Ramanujan graphs\cite{Sarnaknotes, Lubotzky, Lubotzky2019}. They were demonstrated in a celebrated paper by Lubotzky, Phillips and Sarnak\cite{Sarnak1988} using Cayley graphs based on the group $PSL_2(\mathbb{Z}/q\mathbb{Z})$ (the projective special linear group $PSL_n(\mathbb{Z})$ is defined by the induced action of the special linear group on the associated projective space). The method shows an ingenious way to produce edges on all scales using group operations from specially chosen generators. In that work, explicit construction of a family of $d$-regular Cayley graphs with an optimal spectral gap exploited a deep connection to number theory\cite{DSV}. It is this link to the Ramanujan conjecture\cite{Li2019} that gave the graphs their name. Again, for details see Ref. \cite{Lubotzkybook}. 

\begin{Definition}
	A sequence of $d$-regular graphs on $n$ vertices $G(n,d)$ are Ramanujan if for all $n$ and $j = 1, \dots, n-1$ (or $j = 1, \dots, n-2$ if the graph is bipartite) we have
	\begin{equation*}
		\Big| \lambda_i(G[n,d]) \le 2 \sqrt{d - 1} \Big| .
	\end{equation*}
\end{Definition}

Ramanujan graphs are optimal expanders. Peter Sarnak writes\cite{Sarnaknotes}: ``Ramanujan graphs are highly connected sparse large graphs. The tension between sparse and highly connected is what makes them so useful in applications.''

\subsection{Robustness of expander graphs}

An advantage of using $d$-regular expander graphs for QL graph applications is that the states of these graphs are remarkably resilient to disorder in the structure of the graph. To start with, obviously the ensemble of $d$-regular random graphs is already huge and each graph has precisely the same emergent eigenvalue as all others in the ensemble. Moreover, the principle of the construction is simple, in the sense that the only requirement is that each vertex connects to $d$ edges. This simple design rule means that the entropic cost to produce a spectrum containing the ideal emergent state, with state vector $(1, 1, 1, \dots)/\sqrt{n}$, is surprisingly very small---which is evidenced by these states being produced spontaneously in oscillator networks\cite{Townsend1, Townsend2, Strogatz2000, StrogatzBook}. One way to think about this is that the vast number of random states greatly `outweigh', and thereby counter-balance, the cost of producing one perfect state. 

\begin{figure}
	\includegraphics[width=7 cm]{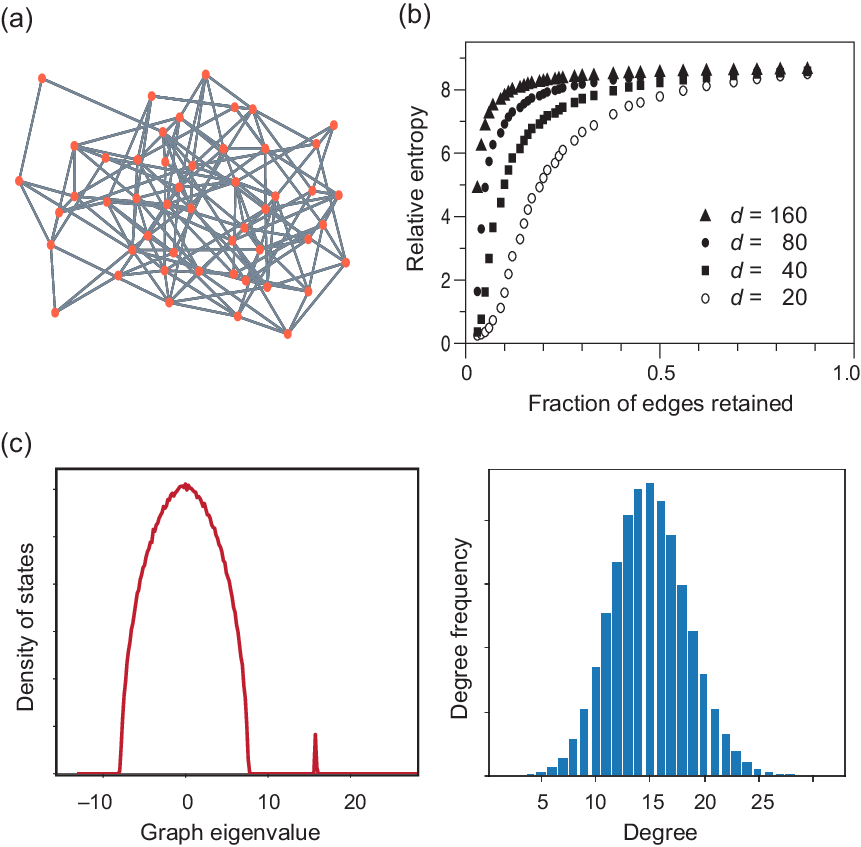}
	\caption{(a) Example of a small $d$-regular random graph with 40\% of the edges removed at random. (b) Purity of the emergent state as a function of the fraction of edges retained in ensembles of $d$-regular random graphs.  (c) Spectrum and analysis of the degree distribution for a $d$-regular random graph ensemble where over half the edges have been randomly removed from each graph. See Ref. \cite{ScholesEntropy} for details. }
	\label{figdisorder}
\end{figure}

It turns out that we can significantly disrupt a perfect $d$-regular random graph and retain a near-perfect emergent state\cite{ScholesEntropy, ScholesQLstates, Scholes2020}. To illustrate this, results of numerical studies from Ref. \cite{ScholesEntropy} are shown in Fig. \ref{figdisorder}. A small disordered $d$-regular graph is drawn in Fig. \ref{figdisorder}a. This is based on a $d = 10$ graph on 50 vertices, but the graph shown has 40\% of the edges removed at random so that it now is more like a $d = 4$ graph, but clearly with a distribution of degrees. This is the kind of construction, for larger graphs, explored in Ref. \cite{ScholesEntropy}. 

In Fig. \ref{figdisorder}b, c we study ensembles of randomly-generated $d$-regular random graphs on 400 vertices. Into each graph we have introduced a number of random edge deletions. For each graph, we then calculate the eigenvectors and eigenvalues. The the eigenvector of the emergent state in each graph in the ensemble was converted to a density matrix (obviously a pure state) and the ensemble density matrix was produced by a convex sum. In Fig. \ref{figdisorder}b the purity of the emergent state of sequences of ensembles of $d$-regular random graphs with various values for $d$ is plotted as a function of the fraction of edges randomly retained after the random deletions. Recall that purity is a gauge of the mixedness of the state\cite{ScholesQIS}. Notice that we can randomly remove about half the edges and still retain an emergent state that is not significantly mixed. 

The robustness of states of these graphs to edge deletion might have some connection to Braess’s Paradox\cite{BParadox}, which is relevant for certain kinds of networks, such as those with flow or game rules. Braess’s Paradox refers to the phenomenon possible in models of heavy traffic flow and similar scenarios where, counter-intuitively, the traffic flow may not be adversely affected by blocking a road. In some cases the flow may even be improved. The principle can be explained in terms of the spectral gap, equivalent to $\lambda_0 - \lambda_1$ in our notation. The spectral gap is related to the reciprocal of the relaxation time for a random walk on the network and it is therefore used as a measure of congestion (for instance in the physical example of a traffic network). Obviously, adding edges to a random network ultimately maximizes the spectral gap by increasing the average vertex degree\cite{ScholesEntropy}. However, adding certain edges---in the interim---can actually decrease the spectral gap, that is, increase congestion\cite{Rudelson2017, BParadoxDeloc}.

In Fig. \ref{figdisorder}c the spectrum of the ensemble of graphs is plotted, in this example the base graphs are $d$-regular random graphs on 400 vertices with $d = 80$ and merely $3/16$ of the edges from the base graph are retained. The semicircle of random states are evident along with the emergent eigenvalue. In each example studied\cite{ScholesEntropy}, the eigenvalue of the emergent state was found, not at $d = 80$, but at $d'$, where $d'$ is the average degree of all vertices in the graph. A histogram of the distribution of vertex degrees throughout the ensemble of graphs is shown. It appears that the average degree is the key parameter, even though there are many vertices with less edges, as seen in the distribution. That point is discussed further in Ref. \cite{Excitonics2024}. 

The possibility that the subgraphs used in the QL construction can be quite disordered means that there is a low `cost' to generating them---they do not need to be precisely engineered. Therefore, it is reasonable to imagine that they exemplify a resource that could evolve naturally in complex systems, including biological systems. Indeed, it is difficult \emph{not} to generate an expander graph when generating a random network with sufficient edges, particularly if there is a predisposition to distribute edges more-or-less evenly among the vertices.

\subsection{Summary: Why use expander graphs?}

Why base the QL bits and states on $d$-regular random graphs and not simply a network of coupled oscillators? There are several advantages to using the expander graphs, which include the following. (i) The expander graph `object' allows us to define measurement projections in terms of edge bias rather than physical orientation. (ii) Expander graphs are robust and easy to construct in any complex system. Moreover, which ever way we construct the graph, in terms of number of vertices and precise edge structure, it always has an eigenvalue of $|d|$. (iii) The spectral gap (existence of an emergent state) is guaranteed, even in the `continuum limit' of vertices. (iv) The graph model allows physical interpretation as a system of coupled phase oscillators, emphasizing the importance of globally sychronized phases rather than a deterministic Hamiltonian. (v) The underlying group structure is not Abelian (i.e. it is non-commutative) because Abelian groups do not have Kazhdan's property ($T$). (vi) We have the convenience that a product of a $d$-regular random graph is also a $d$-regular random graph.

\section{Example: synchronizable networks}

Any graph, including those that encode QL states, have an obvious interpretation as networks. The vertices of the graph are the nodes of a network---that is, the objects being connected---while the edges define how the objects are connected to each other. Networks of \emph{phase oscillators} are especially appealing for providing physical insight into QL states because they focus on oscillator phase. The collective phase topology of the system is the key to producing QL states. The oscillators can be genuine oscillators, like pendulum clocks, or they can be concentrations of chemical reagents, or a range of other abstract concepts\cite{StrogatzBook}. We have examined this topic in prior work\cite{QLsync}. Note that the map from network to graph need not be one-to-one. For example, many nodes in the network can be mapped to one vertex in the graph by coarse-graining. 

In the phase oscillator model, each vertex can be thought of as an oscillator endowed with a frequency, that we treat in the rotating frame of the network of oscillators by its difference from the mean frequency $\epsilon_i$, and a phase offset $\phi_i$ for the oscillator at vertex $i$. We collect these terms as the time-dependent phase for each oscillator $\theta_i = \epsilon_it + \phi_i$, which is associated with the set of vertices. The oscillators are coupled according to the edges in the graph which, under appropriate conditions\cite{Strogatz2000, Acebron2005, daFonseca2018, Rodrigues2016, ScholesAbsSync}, allows the oscillators to synchronize after several periods of time. The phases associated to the vertices come into play as phase differences $e^{i(\theta_i -\theta_j)}$ multiplying the non-zero entries off-diagonal of the graph's adjacency matrix. This is accomplished by a suitable unitary transformation\cite{QLsync} (specified below). The phases then evolve, according to the Kuramoto model\cite{daFonseca2018, Strogatz2000}, as:
\begin{equation}
	\dot{\theta}_i   = \epsilon_i - \frac{K}{N} \sum_{j=1}^{N} a_{ij} \sin(\theta_j - \theta_i),
\end{equation}
where $\epsilon_i $ is the frequency offset from the mean of the oscillator at node $i$, $\theta_i$ is the oscillator phase defined in terms of accumulated phase and an offset $\theta_i(t) = \epsilon_it + \phi_i$. $K$ is the coupling value, $a_{ij}$ are entries from the adjacency matrix of the graph. The nonlinearity comes from the coupling, which favors minimization of the phase differences.

In a recent study we simulated a phase oscillator network with a coupling structure templated by the Cartesian product of two QL bit graphs $G_A \Box G_B$. The phase oscillators are indicated by vertices of the graph, and are assigned a phase that biases the edges (in addition to the edge bias topology already included in the adjacency matrix). The phases evolve with time according to the Kuramoto model. Each oscillator labeled $j$ is associated with a vertex $j$ and therefore a phase $\theta_j$. The adjacency matrix $A$ contains the fundamental phase structure defined by the graph edges and their biases. It is transformed by the oscillator phases according to:
\begin{equation}
	A^{\prime} = \Phi^{-1}A\Phi ,
\end{equation}
with the unitary matrix $\Phi$ being
\begin{equation}
	\Phi = 
	\begin{bmatrix}
		e^{i\theta_1} & 0 & 0 & \dots \\
		0 & e^{i\theta_2} & 0 & \dots \\
		0 & 0 & \ddots & 0 \\
		0 & \dots & 0 & e^{i\theta_{2n}} 
	\end{bmatrix} .
\end{equation}
This unitary transformation of $A$ does not change the spectrum, of course, but it does rotate the basis for the eigenstates. Specifically, the transformation multiplies edge biases by oscillator phase differences, $e^{i(\theta_j - \theta_i)}$. So, when all the oscillators are in phase, there is no rotation of $A$.

The reasoning motivating that study is that a classical system should have a clearly defined phase topology of some kind in order to produce QL states. The Kuramoto model is sufficiently general that it allows us to assess the emergent states as a function of synchronization of a large phase oscillator network. The model can represent many different physical, biological, social, etc scenarios. 

\begin{figure*}
	\includegraphics[width=8 cm]{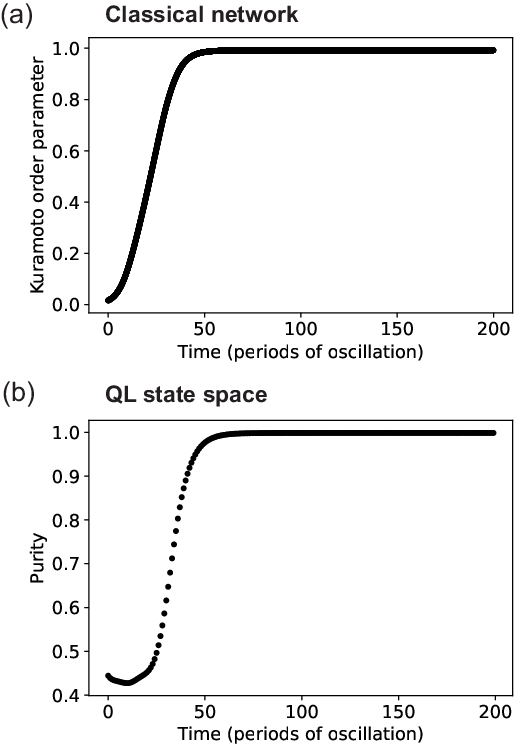}
	\caption{ Numerical results for a network of phase oscillators that are connected in a way prescribed by the product of two QL bits. For these calculations, we  performed an average over initial phases as well as an average over 500 graphs. See Ref. \cite{QLsync} for details. The initial phase distribution was set to be close to the uniform distribution and the coupling that weights the edges was $K = 250$.  (a) Ensemble-averaged Kuramoto order parameter as a function of time predicted for the system of oscillators. The system equilibrates to limit cycle dynamics, indicating a synchronized classical system. (b) The purity of the state associated with the greatest eigenvalue concomitantly increases with time.  Reprinted from G. D. Scholes, 2025, Dynamics in an emergent quantum-like state space generated by a nonlinear classical network, xxx xx:xxxx. arXiv:2501.07500. }
	\label{figSync}
\end{figure*}

At each time point during these dynamics we recorded the density matrix for the state associated with the highest eigenvalue (the emergent state) and accumulated an ensemble average over many realizations of state preparations (the initial phase distribution). During the dynamics, the classical oscillator network synchronizes, Fig. \ref{figSync}a. The purity of the state associated with the greatest eigenvalue concomitantly increases with time, Fig. \ref{figSync}b, showing that the QL state becomes less mixed as the associated classical network becomes more synchronized. 

These results show how the nonlinear dynamics of the classical system translate to a QL state space that is resistant to decoherence\cite{QLsync}. A key finding is that  synchronization of the underlying classical system produces the collective phase structure that, in turn, allows generation of the QL state space. 

\section{Can a classical system generate non-classical correlations?}

\subsection{Nonseparable states of light}

Suitably structured classical light can display \emph{classical entanglement}\cite{nonseplight}. For example, basis states can be polarization, spatial mode, orbital angular momentum (a characteristic of `twisted light'\cite{Barnett-twisted}), etc. Interference of two separable states produced by combining two of these states, each in a superposition, produces classically entangled states of light.

\begin{figure*}
	\includegraphics[width=8 cm]{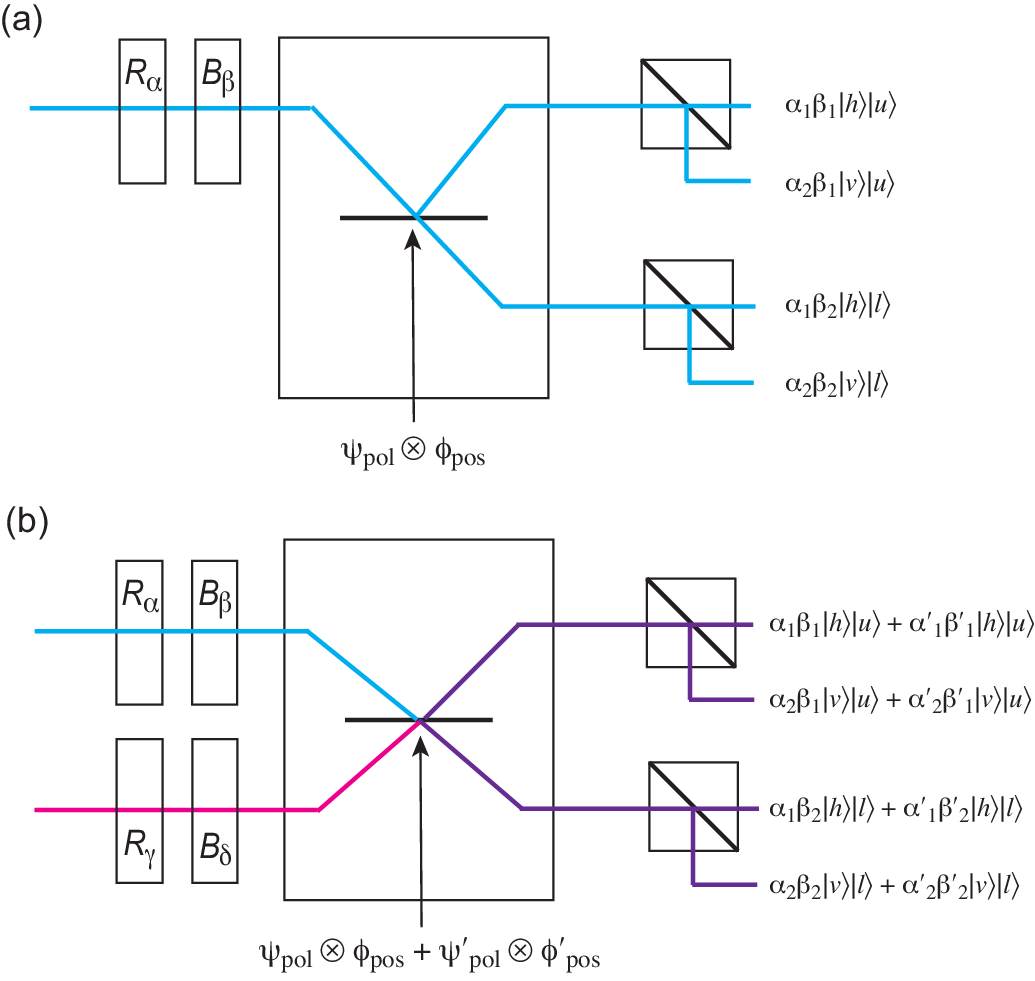}
	\caption{(a) Schematic of an optical setup for producing light classically entangled in position and polarization. Polarization is superposed by the first optical element ($R$), then position is superposed according to the phase conditions set by the second optical element ($B$). $u (l)$ means upper (lower) position. $h (v)$ means horizontal (vertical) polarization. (b) Addition of a second beam can produce a non-separable state by interference at the beam splitter. See Ref. \cite{Spreeuw1998} for details. }
	\label{fig_classicalentangle}
\end{figure*}

Spreeuw\cite{Spreeuw1998} gives a clear example of classical entanglement for light beams, here involving product states of linear polarization and position. The idea is to take as input a pair of light beams and put them into superpositions of product states of polarization and position using a sequence of optical elements, Fig. \ref{fig_classicalentangle}. Here a slightly modified set-up from that shown in the original paper is explained. Two optical elements are used to prepare the state, a polarization rotator ($R$) and a phase rotator ($B$). The fraction of the light beam that is reflected versus transmitted from the beam splitter is controlled by the phase rotator. The classical bits here are polarization ($h$ or $v$) and position ($u$ or $l$) of the beam at the special beam splitter,  denoted $\psi_{\text{pol}} = \alpha_1|h\rangle + \alpha_2|v\rangle$ and $\phi_{\text{pos}} = \beta_1|u\rangle + \beta_2|l\rangle$ respectively.  

In Fig. \ref{fig_classicalentangle}a the generation of a separable state comprising the tensor product $\psi_{\text{pol}} \otimes \phi_{\text{pos}}$ is shown. This state is analogous in concept to the emergent QL states produced by the graph product of two QL bits. The outputs from the measuring device, comprising the beam splitter and polarizers, are presented in the basis of position ($u, l$) and polarization ($v, h$), specifically $|h,u\rangle$, $|v,u\rangle$, $|h,l\rangle$, and $|v,l\rangle$. A rigorous characterization of the state would require a different measuring set-up\cite{James2001}, but the one shown is sufficient for the vertical-horizontal polarization superposition. 

In Fig. \ref{fig_classicalentangle}b, a set-up for producing superposition states, which could potentially include entangled states, is shown. Two different beams are introduced and overlapped on the beam splitter. At the point of overlap, the state of the combined beams is the superposition 
\begin{equation}
	\psi_{\text{pol}} \otimes \phi_{\text{pos}} + \psi^\prime_{\text{pol}} \otimes \phi^\prime_{\text{pos}}.
\end{equation}
The beams overlapping on the beam splitter generate this state by classical interference. 

A correlation coefficient was defined by Spreeuw in terms of intensities measured by the four detectors, $I_{++}$, $I_{+-}$, $I_{-+}$, and $I_{--}$, and unitary transformations that change the basis of the outputs. One can then define a combination of measurements in terms of different measurement angles. Thus a CHSH-type inequality can be tested and it can be confirmed that it is violated with the choice of suitable angles\cite{Spreeuw1998}. This prediction suggests that the correlations in this classical set-up exceed those characteristic of a classical system, just like we expect for an entangled quantum system. However, Spreeuw writes:

\begin{quote}
``It seems that the measured intensities in the classical case can be mapped one-to-one on the coincidence count rates in an EPR type experiment. However, there is one key ingredient missing in the pair of classical light beams: \emph{nonlocality}. Bell's inequality and the experiments testing it lose their significance without nonlocality.''
\end{quote}

The fundamental concern that has been raised is that one of the cebits (Spreeuw's QL bits) cannot be separated from the other, which is the issue with classical entanglement already noted. 

Paneru and co-workers\cite{Paneru2020} give a survey of entanglement versus `classical entanglement'. They argue that ``classical correlations cannot lead to the same conclusions as quantum entanglement''. They note the importance of being able to separate spatially the qubits (QL bits in our case). We address this particular issue below. Paneru and co-workers\cite{Paneru2020} make the important observation that quantum systems are comprised of particles, like photons or electrons and so on, that display wave-particle duality. This is indeed a crucial difference between quantum systems and QL systems generally. 

However, an important difference between quantum superposition states and the non-separable states of classical light described in the example above is in the very nature of the superposition. It is produced in the example by classical interference between the two overlapping light beams to produce a single output state. That output state exists only on the beam splitter. 

The beam splitter is part of the measuring apparatus and subsequently dissects the state into contributions from upper and lower basis states, $\psi_{\text{pol}} \otimes \beta_1 |u\rangle + \psi^\prime_{\text{pol}} \otimes \beta^\prime_1 |u\rangle$ and $\psi_{\text{pol}} \otimes \beta_2 |l\rangle + \psi^\prime_{\text{pol}} \otimes \beta^\prime_2 |l\rangle$, respectively. Owing to the way that classical interference works by adding amplitudes, it can combine states (using the QL state notation from Eqs. 19) to produce non-separable classical light versions of the Bell states as follows:
\begin{subequations}
	\begin{align}
		\Phi_+ = v_{++} + v_{--} = |a_1\rangle |b_1 \rangle + |a_2\rangle |b_2 \rangle  \\
		\Phi_- = v_{-+} + v_{+-} = |a_1\rangle |b_1 \rangle - |a_2\rangle |b_2 \rangle \\
		\Psi_+ = v_{++} - v_{--} = |a_1\rangle |b_2 \rangle + |a_2\rangle |b_1 \rangle \\
		\Psi_- = v_{-+} - v_{+-} = |a_1\rangle |b_2 \rangle - |a_2\rangle |b_1 \rangle .
	\end{align}
\end{subequations}

There are two concerns here. First, these states are produced by combining two separate systems---the two light beams---and are therefore not intrinsic to a single system. Second, combining states in this way is a convex sum operation and should, in the context of quantum information science, produce a statistical mixture of the two states rather than a new pure state.

\subsection{Non-classical correlations in QL states}

Unlike the example discussed in the previous section, QL states are generated by a single system, the QL graph. We could produce a similar interference effect as that described above, but it would require the superposition of emergent states from two \emph{disconnected} graphs. Instead, let's see what is possible from a single QL graph.

The emergent states of QL product graphs are separable states, by construction. So, how can we produce the superpositions needed for entanglement? One solution is to assign frequencies to the subgraphs. For instance, if the frequency associated to $|a_1\rangle$ and $|b_1\rangle$ is $\omega_1$, while to $|a_2\rangle$ and $|b_2\rangle$ we associate a frequency $\omega_2$, then the product states $|a_1\rangle |b_1 \rangle$ and $|a_2\rangle |b_2 \rangle$ can be detuned from $|a_1\rangle |b_2 \rangle$ and $|a_2\rangle |b_1 \rangle$. This produces genuine entangled states, but not maximally entangled states.  A second solution is to control the physical QL network using gate operations inspired from quantum computing. We have explored that approach in some detail in prior work\cite{Amati1, Amati2}. For example, by applying a suitable unitary transformation to the network edge biases we can generate the GHZ state\cite{QLproducts}.

Here we explore how entanglement might emerge naturally from the product structure when QL bits are combined. In particular, we aim to devise a construction that is the QL analog to the generation of non-separable classical light so that we can compare and contrast these concepts. An approach is suggested by the spectrum of the QL state comprising the graph product of two QL bits, Fig. \ref{figSync4}b. Notice that the middle pair of states have the same eigenvalue. Notice also that any pair of the QL states $v_{++}$, $v_{--}$, $v_{+-}$, and $v_{-+}$ can correspond to these degenerate states, simply by setting the edge biases appropriately. When we do this, we do not produce a single state by classical interference, but instead we produce a mixed state---a statistical sum of the two states. For example, let's consider the case when $v_{++}$ and $v_{--}$ have the same eigenvalue. The corresponding mixed state is given by the density matrix:
\begin{equation}
	\rho_{++--} = \frac{1}{4}
		\begin{bmatrix}
	    1 & 0 & 0 & 1 \\
		0 & 1 & 1 & 0 \\
		0 & 1 & 1 & 0 \\
		1 & 0 & 0 & 1 
	\end{bmatrix} .
\end{equation}
This is not the density matrix we would obtain by classical interference, giving the `pure' classical state $\Phi_+ = v_{++} + v_{--} = |a_1\rangle |b_1 \rangle + |a_2\rangle |b_2 \rangle$. This, instead, is a mixed state with concurrence measure of 0.

Then, how do we think about the state $\rho_{++--}$? Recall from linear algebra that when two states share the same eigenvalue, then any convex sum of those eigenstates are also valid eigenstates. This suggests that a better way of thinking about the two states in the mixture would be to consider permutations of the graph vertex sets as operations in the symmetric group algebra, precisely like we do for electrons\cite{Pauncz}.  Irreducible representations of the symmetric group suggest that appropriate eigenvectors in the mixture are $\Phi_+ =  |a_1\rangle |b_1 \rangle + |a_2\rangle |b_2 \rangle$ and $\Psi_+ = |a_1\rangle |b_2 \rangle + |a_2\rangle |b_1 \rangle$. Now we see that $\rho_{++--} = \frac{1}{2}\rho_{\Phi+} + \frac{1}{2}\rho_{\Psi+}$,  indicating that the density matrix is a convex sum with equal weights of these two pure states. The $\Psi_+$ component is notably missing from the state produced by classical interference. 

Notice that the two states in the mixture, $\Phi_+$ and $\Psi_+$, are associated physically on complementary vertex sets. However, we cannot access these states separately simply by disconnecting the product graph to isolate these two vertex sets. The graph only has meaning as an entirety because this is how the product basis is generated. To conclude, mixed states produced by eigenvalue degeneracy in the QL state spectrum can be thought of as statistical mixtures of two maximally entangled states. Nevertheless, to access the entanglement we somehow need to separate those states, or distinguish them by measurement operations, without disconnecting the product graph structure. In the following subsection, an approach for accomplishing this separation of states is devised and discussed. 

Producing the non-separable states using this strategy of exploiting separable states with the same eigenvalue can be accomplished only for central states in the spectrum. This means that the highest (or lowest) eigenvalue emergent state will remain separable. This observation indicates the importance of accessing a spectrum of states to produce non-classical correlations. It also suggests that maximal entanglement is not evidenced in the ground state of classical systems that synchronize according to the emergent state---the system needs to attain a kind of harmonic of that synchronization---unless we apply specially designed gate operations to the network\cite{Amati1, Amati2}.

\subsection{What does nonlocality mean in the context of QL states?}

QL states are generated by classical systems, so does it make sense to think about a QL analog of nonlocality? In particular, the QL system models the correlation space of the states, so is it possible to perform local measurements on the QL bits that were brought together by the Cartesian product? How can appropriate measurements  be performed on the classical system. For example: What does `collapse' signify classically? How can we perform local measurements once we have produced a classical system that represents a product state? This section presents some ideas that can be viewed as a starting point for future work.

To frame the question of measurement and nonlocality, let's consider two QL bit graphs ($G_A$ and $G_B$) and think of them as `atoms', each hosting a QL state vector when they are isolated from one another. We can position these graphs physically in space however we wish. Similarly, we can imagine making a measurement on either QL bit. A measurement in this setting might be accomplished by using a witness for the overall phase so that we can differentiate, in the context of Table 1, states with eigenvalue $d$ from those with eigenvalue $-d$. A suitable witness could be provided by a QL bit (we label it $X$) with known overall phase, such as the $x$-projection where the emergent state is $|x_1\rangle + |x_2\rangle$. Now couple the QL bit $X$ to our QL bit of unknown state and allow the phases to synchronize. The overall phase of $X$ will be unchanged by coupling to another QL bit with the same overall phase, that is, whose $x$-projection of the emergent state is $|a_1\rangle + |a_2\rangle$. However, the phase will invert if coupled to a state where the $x$-projection of the emergent state is $|a_1\rangle - |a_2\rangle$. The reasoning is discussed in Ref. \cite{ScholesQLstates}.

The state space of the entangled pair of QL bits, though, comes from the product graph $G_A \Box G_B$. But now we have completely lost any notion of separate QL bits that we can position in space or measure independently.  Therefore it is difficult to decide how nonlocality can be manifest. 

To emphasize why the product itself is not enough to define nonlocality, we can recall that there is no relevant notion of distance between QL bits because of the definition of the tensor product (and graph product). For example, considering the state space, there is no concept of distance between states in $\mathcal{H}_A$ and those in $\mathcal{H}_B$ in the Hilbert space $\mathcal{H}_A \otimes \mathcal{H}_B$. That is because the basis vectors in $\mathcal{H}_A \otimes \mathcal{H}_B$ are entirely new vectors compared to vectors in either $\mathcal{H}_A$ or $\mathcal{H}_B$. There is no inherent distance function in the tensor product space, as is well known, but we can define a distance by leveraging the inner products in the constituent Hilbert spaces. We use Theorem 6.3.1 in \cite{Murphy}:
\begin{Definition}
	Let $\mathcal{H}$ and $\mathcal{K}$ be Hilbert spaces with $u, v, \in \mathcal{H}$ and $x, y, \in \mathcal{K}$. Then there is a unique inner product $\langle \cdot, \cdot \rangle$ on $\mathcal{H} \otimes \mathcal{K}$ such that 
	\begin{equation*}
		\langle u \otimes x, v \otimes y \rangle =  \langle u, v \rangle  \langle x, y\rangle.
	\end{equation*}
\end{Definition}

Recalling that the norm for a vector in a Hilbert space is provided by an inner product, we can compute $\| u \otimes x - v \otimes y \|$ for $u, v \in \mathcal{H}_A$ and $x, y \in \mathcal{H}_B$. One thus sees that the distance between the vectors is given in terms of products of inner products \emph{within} each Hilbert space, not \emph{between} Hilbert spaces. This result is evident also in the graph product $G_A \Box G_B$. If we construct the product by copying $G_A$ at every vertex of $G_B$, then connect these graphs with the edges of $G_B$ as in Definition 2. Now, in $G_A \Box G_B$ we find only edges transcribed from $G_A$ and edges from $G_B$---there are no edges \emph{between} $G_A$ and $G_B$.

Nonlocality would be evidenced by objects allowing us to `read out' from that correlation space. Specifically, we need to be able to address each QL bit in the product as if they were entangled atoms. We have shown that this is impossible using the QL product graph that emulates the quantum state space. This is because the QL product graph is a physical mapping of the state space, no longer related closely to the QL graph `atoms' that we imagined to generate the space by the graph product. We need somehow to append this product graph with QL bit graphs that provide windows to the many repeated copies of each graph in the Cartesian product of QL bit graphs. 

\begin{figure*}
	\includegraphics[width=8 cm]{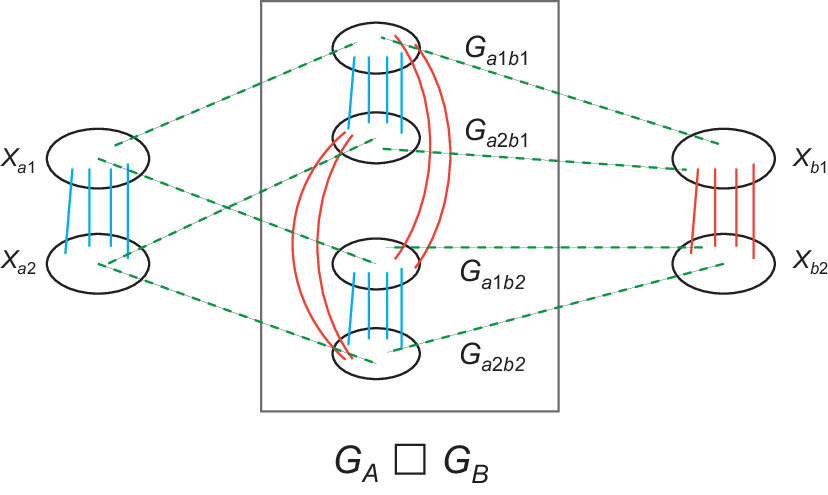}
	\caption{Proposal for appending a QL product graph with witness QL bits that enable read-out or control of QL bits in the product.}
	\label{fig-nonlocal}
\end{figure*}

A proposed strategy is shown in Fig. \ref{fig-nonlocal}. The idea is that a copy of each QL bit graph---we refer to these as witness QL bits designated $X$---composed in the product is connected to the product graph in such a way that  subgraphs in a witness graph are connected to like subgraphs in the product. For example, notice that subgraphs $X_{a1}$ link to all subgraphs $G_{a1*}$, where * means any subgraph of the QL bit labeled B. If there are $q$ QL bits in the product, then each subgraph of a witness QL bit has $k\frac{1}{2}2^q$ edges connecting to the product graph, where $k$ is the number of edges chosen to link witness subgraphs to graph product subgraphs (the dashed lines in Fig. \ref{fig-nonlocal}). Notice that as the product graph becomes larger, each witness graph couples to more and more subgraphs, which will tend to force the witness to synchronize to the QL bit states of the product graph.

The edges from the witness QL bits to the product graph can be made `weak' by setting their entries in the adjacency matrix small. Or, conversely, they can be made `strong'. Strong edges (that is, edges with large bias or coupling strength) could allow the witness QL bits to control the corresponding QL bits embedded in the product, which might be a way to perform gate operations.  Weak edges  allow the witness QL bits to copy the state of corresponding QL bits in the product without significantly perturbing the QL graph. This could enable measurements to be performed on any of the QL bits in the product.  

The proposed structure has an important feature---the QL bits that give windows into the QL correlation space are spatially separable. Does this mean we have endowed our entangled QL system with the extra property of QL nonlocality? The question remains open for now, but the likely answer is that this \emph{QL nonlocality} is not the same as nonlocality in the quantum world, and it needs to be carefully assessed. Yet, it may provide advantages for mimicking additional features of quantum technologies using QL systems. 

\begin{figure*}
	\includegraphics[width=8 cm]{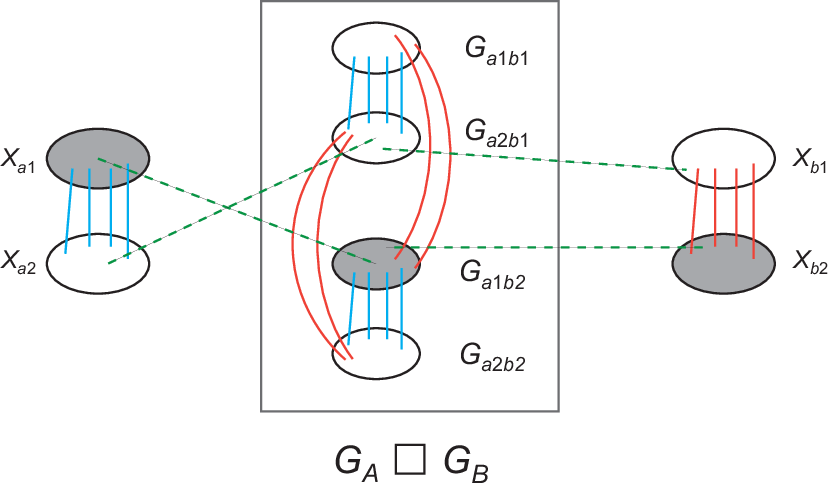}
	\caption{Protocol for selected read-out of the maximally entangled state $\Psi_+$ from the mixed state $\rho_{++--}$. See text for explanation. }
	\label{fig12-nonlocal}
\end{figure*}

The witness QL bits can be used to study the entangled states in the mixture of states discussed in the previous subsection. For instance, $\Psi_+ = |a_1\rangle |b_2 \rangle + |a_2\rangle |b_1 \rangle$ can be detected independently of $\Phi_+$ by connecting $X_A$ and $X_B$ only to the subgraphs corresponding to the appropriate basis states, as shown in Fig. \ref{fig12-nonlocal}. Thus we retain the entire correlation space---the entire product graph---but we can project the entangled state of interest onto the witness QL bits. If we then measure in the $z$-basis (see Table 1) that $X_A$ is in the state $|a_1\rangle $, shown by the shaded subgraph, then by propagating the subgraph with connected vertices, shaded subgraphs, we see that a measurement on $X_B$ must detect $|a_2\rangle $. This is obviously a classical correlation, but notice that the apparent instantaneous response of $X_B$ to measurement of $X_A$ is a kind of illusion that comes about because each QL bit witness reads out from a common correlation space, defined by $G_A \Box G_B$.

\section{Outlook}

We have focused on the way a suitably designed classical system can generate a representation of a quantum-like (QL) state space.  In general, classical systems cannot be associated with a QL state space, but the present work shows of the existence of such maps. The representation is accomplished using a special graph topology.  The graph defines the network structure of the classical system and, in the `other direction', it defines an associated vector space. In particular, we developed a graph topology that produces arbitrary superpositions of states in a tensor product basis. The work highlights the importance of phase coherence in the classical system.

We showed here how the graph product can be optimized to produce a more compact graph with the essential properties required to generate states that mimic many of the properties of quantum states.  This optimized product representation gives a concrete visualization of the correlation structure in a quantum state space that produces important properties like entanglement. We also proposed a way of measuring the states of QL bits that are hidden in the Cartesian product graph. The opens possibilities for realizing classical systems where there is potential for an entanglement detection protocol.  We conclude by suggesting some open questions. 

The graphs allow interpretation of quantum correlations and suggest ways to think about nonlocality. This highlights questions. For example, what does it mean to measure a \emph{state} at a location in space? In particular, what does such a measurement mean for a state defined in a tensor product of Hilbert spaces? Similarly, if we construct a physical representation of a QL graph, how can we measure the properties of one of the QL bits, or its state, once it is incorporated in the Cartesian product? The solution appears to require appending `atomic graphs' (that we call witness graphs) to the product graph. These `atoms' allow suitable read-out from the Cartesian product graph.

Demonstrating how to exploit the QL states for computational or functional advantage is an open challenge\cite{Amati1}. Some points are worth noting. Regardless of QL phenomena, the underlying classical network incorporates intrinsic parallelism, like the universal memcomputing machines envisioned by Di Ventura and co-workers\cite{DiVentura1}. These machines have have the capacity to solve NP-complete problems in polynomial time\cite{DiVentura2}. Coupled oscillator networks are a compelling example of a system that can be realized as a QL resource, and such systems have already been developed for computing\cite{Csaba2020}. These platforms could certainly be adapted to network architecture designed to produce QL states. All these known examples use the classical network for computation. The realization of a map to a QL state space opens up the possibility that the emergent states can also be used as a computational resource, which could allow classical circuits to perform calculations in a similar way to quantum computers, by using superpositions of states\cite{Amati1}.

Further relating to this topic of exploiting the QL graph product structure for functional advantage. Can the  structure of basis states generated by the graph product be used to give advantages for function or computation? Could such function be feasible in biological or soft-matter systems by exploiting manifestation of phase topology and scale-free connections? Or even for QL decision making\cite{Pothos2022, Khrennikov2023book, Ozawa2020, Busemeyer2014}?

The development of experiments to probe some of these questions will be important. For instance, experiments that compare measurements on the space of physical objects compared to the state space they produce in the context of QL states. This could be achieved by studying suitably designed networks\cite{QLsync}. Similarly, physical examples of classical systems templated by QL graphs could allow systematic exploration of measurements on the state space, potentially enabling new ways to examine quantum theory and quantum correlations. In addition, it will likely be interesting examine continuum limits (in the vertices) for the graphs. For example, it might be possible to find a QL uncertainty relation.

\section{Appendix: Symmetric and alternating tensors}

So far we have produced product states, but have not worried about the ordering of the states in the tensor product. We may, however, wish to write linear combinations of tensor product states that satisfy certain permutation symmetries, like bosons and fermions. We can account for the properties of permutations by adapting methods described in detail in \cite{TensorSpaces}.

Let $V$ be our vector space, which has dimension 2 for the QL bit basis we are using. We are interested in the tensor space
\begin{equation*}
	T^n(V) = \underbrace{ V \otimes \dots \otimes V }_{\text{\emph{n} times}} ,
\end{equation*}
which has dimension $2^n$. We can count the number of `2-states', that is, $|a_2\rangle$, $|b_2\rangle$, etc., in the tensor product and designate it $p$, which takes the values 0 to $n$. For each $p$, we have $\binom{n}{p}$ tensor products, and in total $\sum_{p=0}^n \binom{n}{p} = 2^n$.

The entire space $T^n(V)$ comprises two distinct subspaces: the subspace of symmetric tensor product states and that of antisymmetric, or \emph{alternating} states. Let $\mathscr{S}_n$ be the set of permutations of the $n$ states in the tensor product. For an introduction to the symmetric group see \cite{Sagan}. Define the signature of a permutation to be $\text{sgn}(\sigma) = 1$ if $\sigma$ is an even permutation and $\text{sgn}(\sigma) = -1$ if $\sigma$ is an odd permutation. The permutations are the unique linear transformations $P_{\sigma}$ of $T^n(V)$ such that for $v_i \in V$:
\begin{equation*}
	P_{\sigma}(v_1 \otimes \dots \otimes v_n) = v_{\sigma^{-1}(1)} \otimes \dots \otimes v_{\sigma^{-1}(n)} .
\end{equation*}

\begin{Definition}
	Let $t \in T^n(V)$. If $P_{\sigma}(t) = t$ for all $\sigma \in \mathscr{S}_n$, then $t$ is called a \emph{symmetric tensor}. If $P_{\sigma}(t) = \text{sgn}(\sigma)t$ for all $\sigma \in \mathscr{S}_n$, then $t$ is called an \emph{alternating tensor} (or antisymmetric tensor). The set of symmetric tensors and the set of alternating tensors are vector subspaces of $T^n(V)$ called $S^n(V)$ and $P^n(V)$ respectively.
\end{Definition}

We define linear transformations of $S^n(V)$ and $P^n(V)$ as follows:
\begin{flalign}
	&\mathsf{S}_n = \frac{1}{n!} \sum_{\sigma \in \mathscr{S}_n} P_{\sigma} \\
	&\mathsf{A}_n = \frac{1}{n!} \sum_{\sigma \in \mathscr{S}_n} \text{sgn}(\sigma) P_{\sigma} ,
\end{flalign}
where the mappings $\mathsf{S}_n$ and $\mathsf{A}_n$ are called the symmetrizer and alternator on $T^n(V)$.

The QL states generated by the Cartesian product are the set of permutations generated by $\sigma_n$, but without distinguishing the quantum particles (e.g. electrons). That is, the product gives explicitly a single (arbitrary) ordering of states in the product basis. With the formalism just described, it is straightforward to adapt the basic QL states to generate basis states in the symmetric or antisymmetric tensor product state subspaces.

\begin{acknowledgments}
This research was funded by the Gordon and Betty Moore Foundation through Grant GBMF7114. Graziano Amati and Debadrita Saha are thanked for comments on the manuscript. Discussions with Noga Alon and Peter Sarnak have been very helpful. Thank you to Brontë Scholes for the cockatoo drawing.
\end{acknowledgments}


\vspace{6pt} 




\bibliography{Scholes_bib_Oct2024b}

\end{document}